\documentclass[a4paper,12pt]{article}

\usepackage{graphicx,epsfig,natbib}

\usepackage{amssymb}

\newcommand{\ie}{{\it i.e.}{\ }}
\newcommand{\eg}{{\it e.g.}{\ }}

\newcommand{\adsurl}[1]{http://adsabs.harvard.edu/abs/#1}%{ADS}}
\newcommand{\doiurl}[1]{http://dx.doi.org/#1}%{DOI}}

\begin{document}

\
\\

\begin{center}
\LARGE Periodic behaviour of coronal mass ejections, \\ eruptive events, and solar activity proxies during \\ solar cycles 23 and 24
\end{center}

\

\centerline{Tatiana Barlyaeva$^1$,
Julien Wojak$^{1,2}$, Philippe Lamy$^{1,2}$, Brice Boclet$^{1,2}$, Imre Toth$^3$}

\begin{center}
 \it\small
$^1$
Laboratoire d'Astrophysique de Marseille, UMR 7326,
%\\ \it\small
 CNRS \& Aix-Marseille Universit\'e,
\\
38 rue Fr\'ed\'eric Joliot-Curie, 13388 Marseille cedex 13, France
\\
$^2$ \it\small
Laboratoire Atmosph\`eres, Milieux et Observations Spatiales, CNRS \& Universit\'e de Versailles Saint-Quentin-en-Yvelines,
%\\ \it\small
11 Bd d'Alembert, 78280 Guyancourt, France
\\
$^3$ \it\small
Konkoly Observatory, Research Centre for Astronomy and Earth Sciences,
\\
Hungarian Academy of Sciences,
%\\ \it\small
H-1121 Budapest, Konkoly Thege Mikl\'os \'ut 15-17, Hungary
\end{center}

\

\begin{abstract}

We report on the parallel analysis of the periodic behaviour of coronal mass ejections (CMEs) based on 21 years [1996 -- 2016] of observations with the SOHO/LASCO--C2 coronagraph, solar flares, prominences, and several proxies of solar activity.
We consider values of the rates globally and whenever possible, distinguish solar hemispheres and solar cycles 23 and 24.
Periodicities are investigated using both frequency (periodogram) and time-frequency (wavelet) analysis.
We find that these different processes, in addition to following the $\approx$11-year Solar Cycle, exhibit diverse statistically significant oscillations with properties common to all solar, coronal, and heliospheric processes: variable periodicity, intermittence, asymmetric development in the northern and southern solar hemispheres, and largest amplitudes during the maximum phase of solar cycles, being more pronounced during solar cycle 23 than the weaker cycle 24.
However, our analysis reveals an extremely complex and diverse situation.
For instance, there exists very limited commonality for periods of less than one year.
The few exceptions are the periods of 3.1--3.2 months found in the global occurrence rates of CMEs and in the sunspot area (SSA) and those of 5.9--6.1 months found in the northern hemisphere.
Mid-range periods of $\approx$1 and $\approx$2 years are more wide spread among the studied processes, but exhibit a very distinct behaviour with the first one being present only in the northern hemisphere and the second one only in the southern hemisphere.
These periodic behaviours likely results from the complexity of the underlying physical processes, prominently the emergence of magnetic flux.

\end{abstract}

\newpage

%=====================================================================================================================================
\section{Introduction}\label{Intro}
%=====================================================================================================================================

Solar eruptive phenomena such as flares, prominences, and coronal mass ejections (CMEs) are very energetic events which can significantly influence the interplanetary environment and space weather conditions \citep{Chen2011,Webb2012}.
Characterizing their temporal evolution and, in particular, detecting possible periodic patterns can contribute to the understanding of the interactions at work and clarify their relationships and ultimately their physical origins.

Quasi-periodic variations have been found in essentially all physical indicators of solar activity extending from the 27-day synodic rotation period to the $\approx$11-year Schwabe Solar Cycle.
Best examples are:
i) the 154-day periodicity found in the temporal distribution of flares \citep{Rieger1984} and subsequently in a variety of solar and interplanetary data
 \citep[and references therein]{Richardson2005}, and
ii) the 1.3-year periodicity detected at the base of the solar convection zone \citep{Howe2000,Howe2007} and in sunspot area (SSA) and sunspot number (SSN) time series  \citep{Krivova2002}.
These multiple periodicities collectively known as intermediate or mid-term quasi-periodicities together with those in the range of 0.6--4 years are often referred to as quasi-biennial oscillations (QBOs) and have been the subject of an in-depth review by \cite{Bazilevskaya2014}.
\cite{Barlyaeva2015} have recently shown that the radiance of the corona exhibits such QBOs sharing the same properties as those resulting from solar activity.

It has been proposed that these periodicities are in one way or the other related to the emergence of magnetic flux from the convection zone \citep{Carbonell1992,Ichimoto1985}.
Since for instance sunspot area, flares, erupting prominences, and coronal mass ejections are all some manifestation of this emergence -- although their mutual relationships are not fully understood -- it is conceivable that they all exhibit the same periodicities.
The case of CMEs has only been recently considered since the continuous observations performed by the {\it Large Angle and Spectrometric Coronagraph} (LASCO; \cite{Brueckner1995}) onboard the {\it Solar and Heliospheric Observatory} (SOHO) since January 1996 offer the most appropriate source to investigate this question.
\cite{Lou2003} examined the first four years of data around the peak of solar cycle (hereafter abbreviated to SC) 23 and based on Fourier power spectral analysis, they found significant power peaks at ten periods ranging from 33.5 to 358 days.
Six of them exceeded two months, namely 66.25 days (2.2 months), 100 months (3.3 months), 110.8 days (3.64 months), 196 days (6.44 months), 272 days (8.9 months), and 358.3 days (11.8 months); note that they exclude the 154-day Rieger period found in flares.
\cite{Lara2008} used the maximum entropy method to compute the power spectrum of a CME time series extending over a time interval of 10.75 years (1996.0--2006.75, that is essentially SC 23) and found ten periods ranging from 17.2 to 408.5 days.
Those which exceed two months are: 93.84 days (3.1 months), 192.9 days (6.34 months), and 408.5 days (1.1 year) and also exclude the 154-day Rieger period.
They also performed a time-frequency wavelet analysis in order to find when the different periodicities took place along the solar cycle.
Contrary to these two studies based on occurrence rates, \cite{Vourlidas2010} investigated the mass rate (as a more relevant physical property)
over a time interval of thirteen years (1996--2009) and applied the Lomb-Scargle spectral analysis to uncover the presence of a 6-month periodicity in the ejected mass from 2003 onward.
In a subsequent erratum, \cite{Vourlidas2011} recognized an error in their previous analysis (failing to take into account the 180$^\circ$ periodic rolls of the SOHO spacecraft) and their re-analysis led to the disappearance of the 6-month periodicity.
They did mention evidences of periodicity but gave no detail in their erratum.
\cite{Choudhary2014} applied standard Fourier analysis to nearly six years (1999.25--2005.0) of CME occurrence rate resulting in a single period of 190 days (6.24 months) and wavelet power spectral analysis (also to flare and sunspot area time series) over a longer time interval of 13 years (1999.0--2012.0) that produced a significant time-frequency area peaking at 193 days and an additional period of about 154 days ($\approx$ 5 months) during the rising phase of the current SC 24.
Their claim that their 6-month period ``is consistent with the findings of \cite{Vourlidas2010,Vourlidas2011}'' is somewhat surprising in view of the retraction by \cite{Vourlidas2011}.
\cite{Guedes2015} used wavelet analysis to identify patterns in CMEs, X-ray solar flares, and SSN in the interval [2000 -- 2013].
The authors found a set of periods in the range of 16--1024 days in CME and X-ray flares appearing and disappearing at different phases of the solar cycle and an additional range of 128--256 days during the rising phase of SC 24 broadly consistent with the results of \cite{Choudhary2014}.

The first aim of this work is to ascertain the existence of periodicities or quasi-periodicities in the CME activity by using a different database than used in the above articles and over a much longer time interval (almost two solar cycles), further evaluating and comparing different techniques of period searching.
We also analyze both the occurrence and mass rates whereas past articles consider only the former rate except that of \cite{Vourlidas2010} which considers the mass but finally did not produce any result.
The second aim consists in comparing the periodicities with those found in the temporal variations of different proxies of solar activity and of erupting processes known to be closely associated with CME, namely solar flares and prominences.
Whereas the understanding of the origin of periodicities is presently out of reach as we shall later discuss, we may hope to shed some light on the underlying process(es) by comparing the results for different solar phenomena.

The article is organized as follows.
Section~\ref{Data} presents the ARTEMIS II catalog of CMEs, the solar proxies selected for comparison, and the solar flares and prominences data.
Section~\ref{Meth} describes the methods used for period analysis. In Section~\ref{Per_CME}, we broadly characterize the
temporal evolution of the CME occurrence and mass rates (globally and by
hemispheres), and then analyze in detail their short- and mid-term variations.
Section~\ref{Per_Others} is devoted to the analysis of high- to mid-term frequency oscillations in proxies of solar activity and in the occurrence rates of flares and prominences.
Finally, we discuss our results in Section~\ref{Disc} and summarize them in the conclusion (Section~\ref{Concl}).

%=====================================================================================================================================
\section{Observational Data} \label{Data}
%=====================================================================================================================================

%-------------------------------------------------------------------------------------------------------------------------------------
\subsection{Coronal Mass Ejections: The ARTEMIS II Catalog} \label{Data_CME}
%----------------------------------------------------------------------------------------------------------------------------------

The aforementioned past investigations were all based on the catalog assembled by the Coordinated Data Analysis Workshop (CDAW) Data Center\footnote{http://cdaw.gsfc.nasa.gov/CME\_list/} which relies on visual detection by different operators.
Limitations and biases inherent to this method (\eg the varying cadence of the LASCO observations and arbitrary criteria resulting in the inclusion of many faint events after 2006) have been repeatedly pointed out \citep{Wang2014,Webb2014} and may a-priori question the validity of this catalog for statistical studies and period searching.
Our analysis is based on the ARTEMIS II catalog \citep{Floyd2013} recognized as the most reliable among the different catalogs \citep{Wang2014}
which is, by its very construction, totally immune to the above problems.
Coronal mass ejections are automatically detected on synoptic maps based on their morphological appearance.
The automated method is based on adaptive filtering and segmentation, followed by merging with high-level knowledge and resulted
in the production of the ARTEMIS I catalog \citep{Boursier2009}.
A new generation of high-definition maps later resulted in the present ARTEMIS II catalog \citep{Floyd2013} which presently covers 21 years (1996 to 2016 inclusive), except for a short interruption when the SOHO spacecraft lost its pointing from 25 June to 22 October 1998 with normal operations resuming only in March 1999.

\begin{figure}
\noindent
\centering
\includegraphics[width=\textwidth]{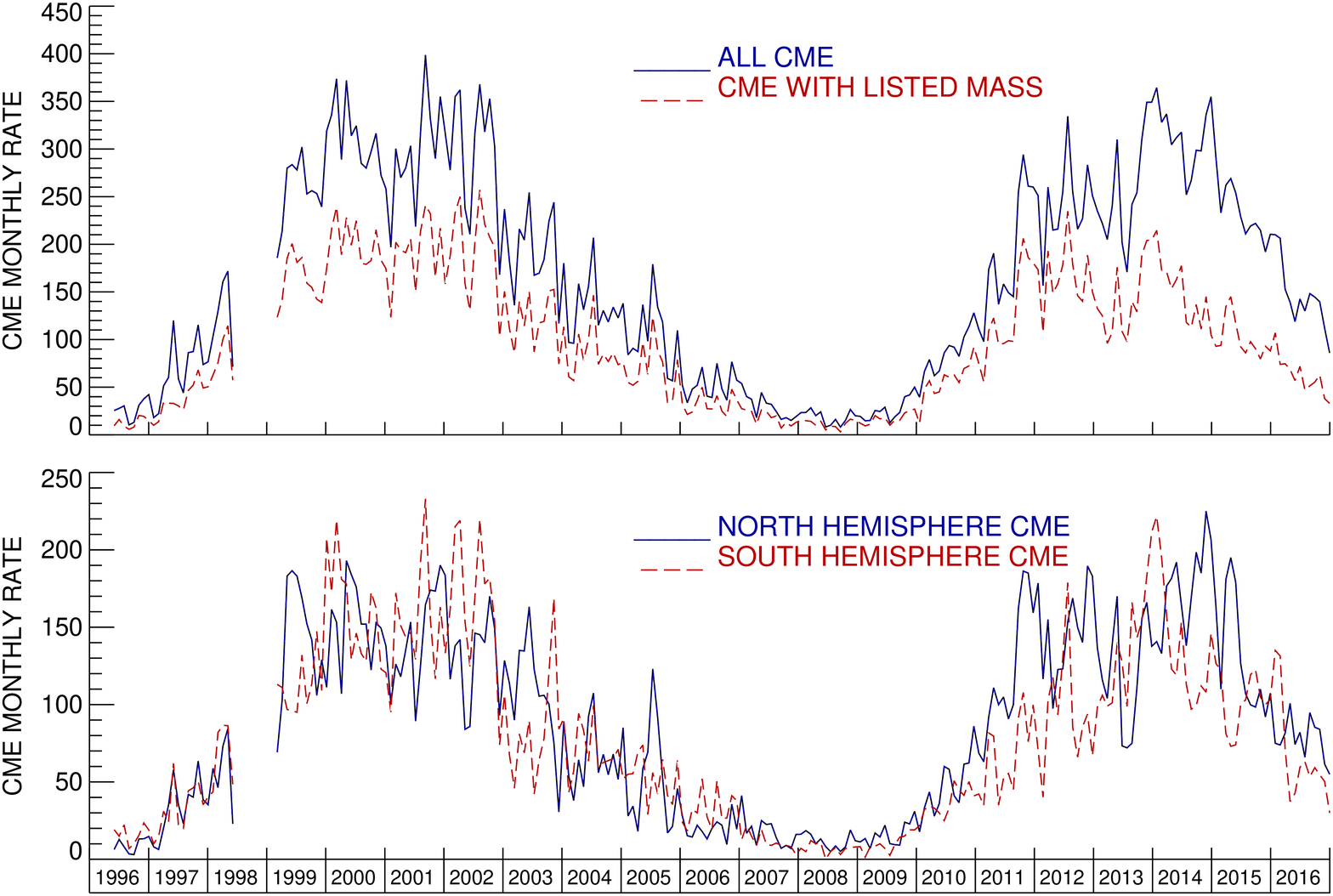}
\caption{Monthly occurrence rates of CMEs (ARTEMIS II catalog). The upper panel displays the case of the global population and of the subgroup of CMEs with a listed mass. The lower panel displays the global population separately in the northern and southern hemispheres.}
\label{fig:plots_CME}
\end{figure}

This global set of CMEs comprises 37790 events, approximately twice the number reported by the CDAW catalog, but comparable to the number reported by the SEEDs catalog\footnote{http://spaceweather.gmu.edu/seeds/}.
The technique used to calculate their mass limits the number to 22468 events ($\approx$60\% of the global population) which defines a sub-set CME$_\textrm{m}$.
We have verified that this selection does not introduce a bias and a visual verification can be performed by inspecting Figure~\ref{fig:plots_CME} which displays the temporal evolution of the monthly occurrence rates.
It can be seen in the top panel that the rate of ARTEMIS II CMEs (blue curve) and that with a mass estimation (red curve) closely track each other.
As a matter of fact, applying a scaling factor of $\approx$1.6 to the latter curve would bring it in almost perfect agreement with the former curve.
Note that, for convenience, our monthly rates are based on a mean month equals to $1/12$ of a year.
We further distinguish the CMEs coming from the northern and southern hemispheres on the basis of their apparent latitude listed in the ARTEMIS II catalog (CME$_\textrm{N}$, CME$_\textrm{m,N}$ and CME$_\textrm{S}$, CME$_\textrm{m,S}$ respectively), and the occurrence rates of the CME$_\textrm{N}$ and CME$_\textrm{S}$ are displayed in the bottom panel of Figure~\ref{fig:plots_CME}.

%-------------------------------------------------------------------------------------------------------------------------------------
\subsection{Description of Selected Solar Proxies}\label{Data_Proxies}
%-------------------------------------------------------------------------------------------------------------------------------------

We consider three photospheric indices: sunspot number (SSN), sunspot area (SSA), and total photospheric magnetic flux (TMF).
The SSN data come from the WDC-SILSO data center\footnote{http://www.sidc.be/silso/datafiles},
and the SSA data from the RGO database\footnote{http://solarscience.msfc.nasa.gov/greenwch.shtml}.
The total photospheric magnetic flux, calculated from the Wilcox Solar Observatory photospheric field maps, was kindly made available to us by Y.-M.~Wang; detail can be found in \cite{Wang2003}.
All indices are considered globally and by hemispheres.

%-------------------------------------------------------------------------------------------------------------------------------------
\subsection{Solar Flares and Prominences}\label{Data_Flares_Protu}
%-------------------------------------------------------------------------------------------------------------------------------------

We consider the GOES flare data from the NGDC/NOAA database\footnote{http://www.ngdc.noaa.gov/stp/space-weather/solar-data/solar-features/solar-flares/x-rays/goes} in different classes depending upon their intensity: A, B, C, M, X.

Prominence data are taken from three different sources.
\begin{itemize}
    \item
    The Nobeyama Radioheliograph (NoRH) database\footnote{http://solar.nro.nao.ac.jp/norh/} \citep{Nakajima1994} based on their detection at 17 GHz.
    We use the so-called ``complete'' list\footnote{http://solar.nro.nao.ac.jp/norh/html/prom\_html\_db/} elaborated to study prominence activity \citep{Shimojo2006,Shimojo2013} and updated to 31 August 2013 (Shimojo, personal communication).
    We restrict the list to eruptive prominences by imposing a minimum height of 1.2 {\it R${}_\odot$} and a positive velocity.
        \item
    The Kislovodsk Observatory database\footnote{http://en.solarstation.ru/sun-service/chromosphere/} based on their detection in H-$\alpha$ \citep{Guseva2007}.
    In line with the above selection, we consider the restricted subset of prominences defined only by their height of at least 1.2 {\it R${}_\odot$} since this
    database does not include velocities.
    \item The catalog of prominence eruptions observed by the {\it Atmospheric Imaging Assembly} (AIA) onboard the {\it Solar Dynamics Obser\-va\-to\-ry} (SDO) and compiled by \cite{McCauley2015}.
    Although limited to slightly more than four years (June 2010 to September 2014), it has the advantage of being free of biases inherent to ground-based observations.
    We consider the restricted subset of unconfined events.
\end{itemize}

Note that the NoRH and Kislovodsk databases use slightly different definitions of the height of the prominences, the former considers their center of mass whereas the latter, the highest part of the prominences.

%=====================================================================================================================================
\section{Methodology} \label{Meth}
%=====================================================================================================================================

In order to investigate periodicities in CMEs, solar proxies and the two eruptive processes, we make use of two different methods: frequency analysis (periodogram) and time-frequency analysis (wavelet).
The former one allows an accurate determination of the periods but lacks temporal information whereas the latter one yields the temporal dependence of frequency ranges whose signal exceeds a given threshold.
In a sense they are complementary and have generally been used in parallel.

Since we are interested in the analysis of high- and mid-range oscillations (from months to a few years), the Schwabe Solar Cycle of solar activity ($\approx$11 years) present in all considered datasets causes a bias which we remove by subtracting a 25-month running average.
This is a standard ``de-trending'' method implemented for instance by \cite{Bazilevskaya2014} in their review of QBOs and we checked that it is indeed an optimal choice for our datasets.

%-------------------------------------------------------------------------------------------------------------------------------------
\subsection{Frequency Analysis: Periodogram} %\label{Meth_1D}
%-------------------------------------------------------------------------------------------------------------------------------------

The most classical tool to detect underlying periodicities in time series is the periodogram which gives a spectral representation of the time series, that is the power of each frequency present in the signal.
It is therefore known as the power spectral density (PSD) whose rigorous definition is the Fourier transform of its autocorrelation.

There is a wide range of methods to perform frequency analysis and a good strategy is to use several of them in order to overcome the drawbacks of each particular one \citep{Berger1991,Pardo2005}.
In this study, we consider the classical methods of Schuster ~\citep{Schuster1898}, Welch ~\citep{Welch1967}, Lomb-Scargle ~\citep{Lomb1976,Scargle1982}, the maximum entropy method \citep{Burg1972} as implemented by \cite{Pardo2005}, several of those having been used in past works on CME periodicities (Section \ref{Intro}).
 We add the window clean estimation ~\citep{Roberts1987} which performs a non--linear deconvolution in the frequency domain (equivalent to a least-squares interpolation in the time domain).
  It improves frequency spectra distorted by the limited frequency resolution due to the finite time span of the data sample and further removes interference oscillations.
The comparison of these methods is presented in Table~\ref{Table_MethP}.
The method of Schuster is an implementation of the discrete Fourier transform and it is equivalent to the Autosignal (v1.7) software used by \cite{Choudhary2014}.
In our case of evenly spaced data (since we consider monthly rates with a mean month equals to $1/12$ year), the Lomb-Scargle method presents no advantage and in fact, produces the same periodograms as the discrete Fourier transform.

\begin{table}[htpb!]
\caption{Comparison of periodogram estimation methods.}
\vspace{0.2cm}
\label{Table_MethP}
\begin{tabular}{cccc}

\hline
                     & Method               & Comments               & Confidence level \\
                     &                      &                        & Type of noise    \\
\hline
Discrete             & Simple and           & Subject to artifacts    & Yes              \\
Fourier transform    & direct computation   & (bias, aliasing)       & Red \& white     \\

\\

Modified Welch       & Mean over            &  No bias              &  No               \\
      periodogram    & windows              & Reduced aliasing      & None              \\

\\

 Lomb-Scargle        & Evaluation on a     & Works with            &  Yes               \\
                     & preset frequency    & uneven sampled        &  White             \\
                     & range               & datasets              &                    \\

\\

Max Entropy          & Estimate             & May give noisy        & Yes                      \\
                     &  Max Entropy         &    peaks              &   Red \& White using      \\
                     & spectrum satisfying  &               &   randomized procedure    \\
                     & variance constraint  &               &                          \\

\\

Window clean         & Joint Fourier       & Produces a           & Comparison          \\
                     & \& deconvolution    & residual             & to residual         \\
                     &  method             & spectra              &  spectra            \\
\hline
\end{tabular}
\end{table}
\newpage
Putative periodicities in time series appear as peaks in periodograms mixed with other peaks coming from fluctuations in the data.
These fluctuations may be real, \ie inherent to the underlying physical processes, or noise resulting from errors in the detection of events and in the measurements of physical quantities.
As far as searching periodicities is concerned, these two spurious sources may be assimilated to and treated as noise.
Therefore, a key aspect of the correct interpretation of the periodograms resides in the definition of a significance level against this noise that will insure the reality of the periodicities with a high probability.
Two approaches are available to solve this question and summarized below; interested readers may consult the quoted references for detailed explanations.

\begin{figure}
\noindent
\centering
CME occurence rate \hspace{2.8cm} CME mass rate\\
\includegraphics[width=0.45\textwidth]{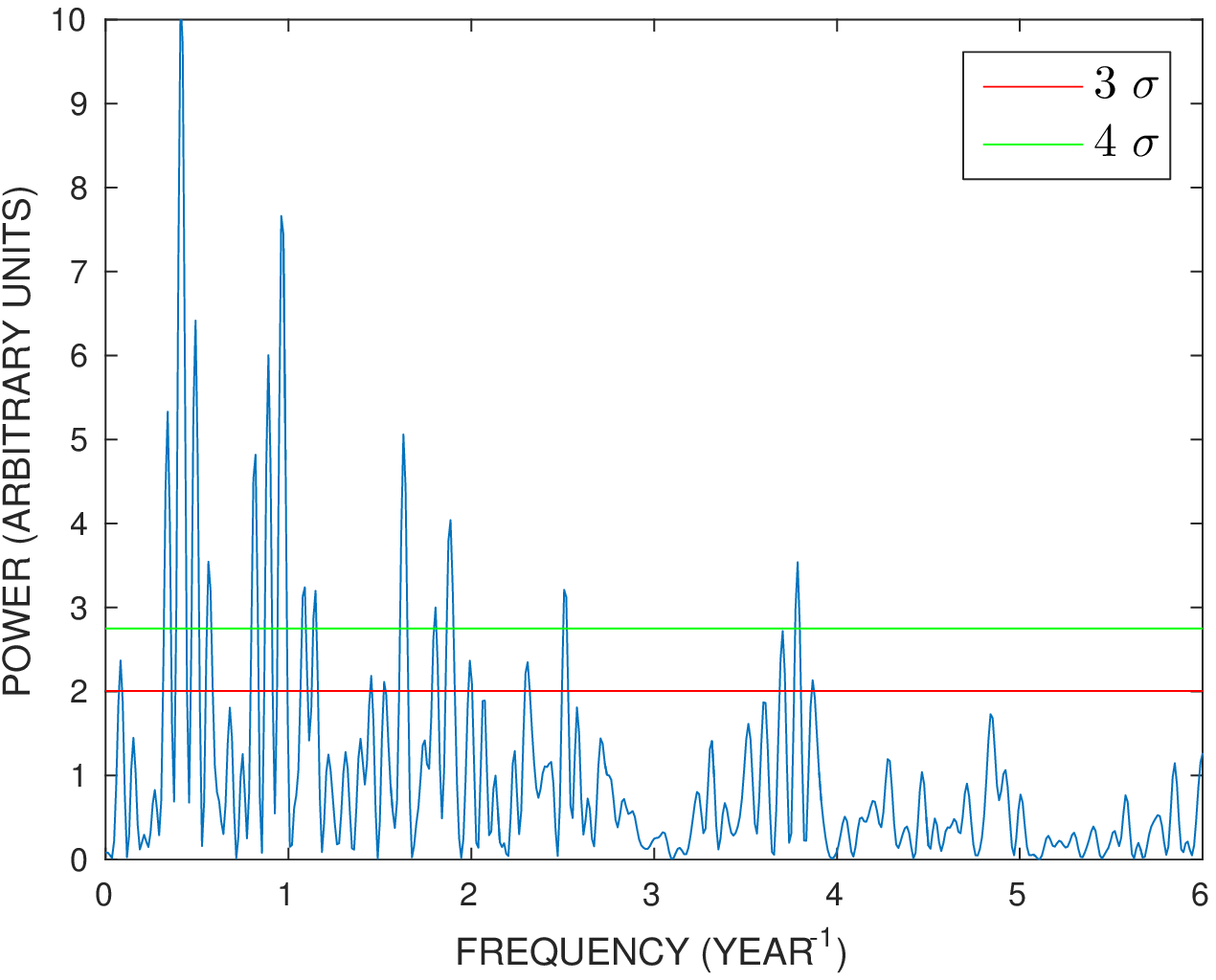}
\includegraphics[width=0.45\textwidth]{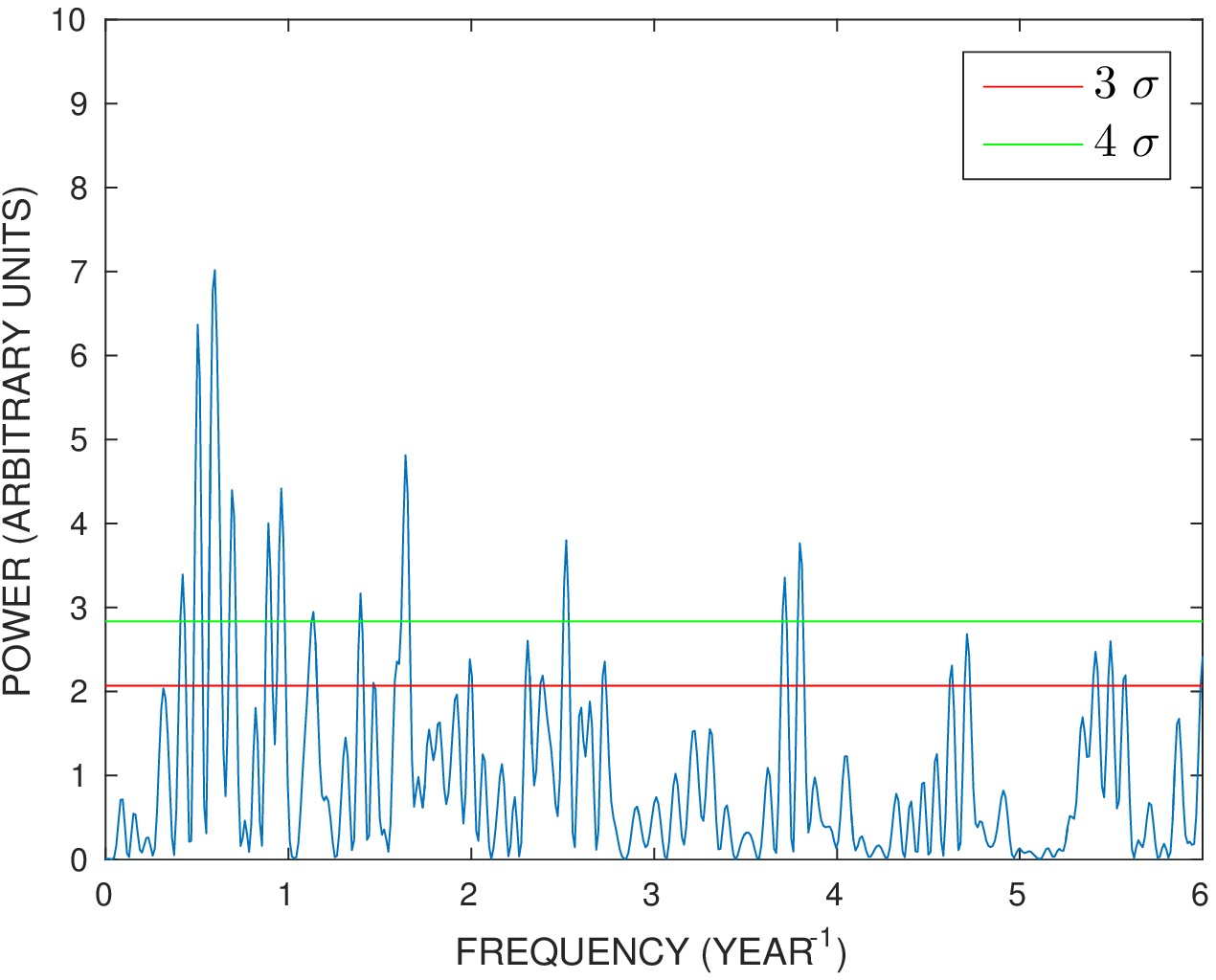}
\includegraphics[width=0.45\textwidth]{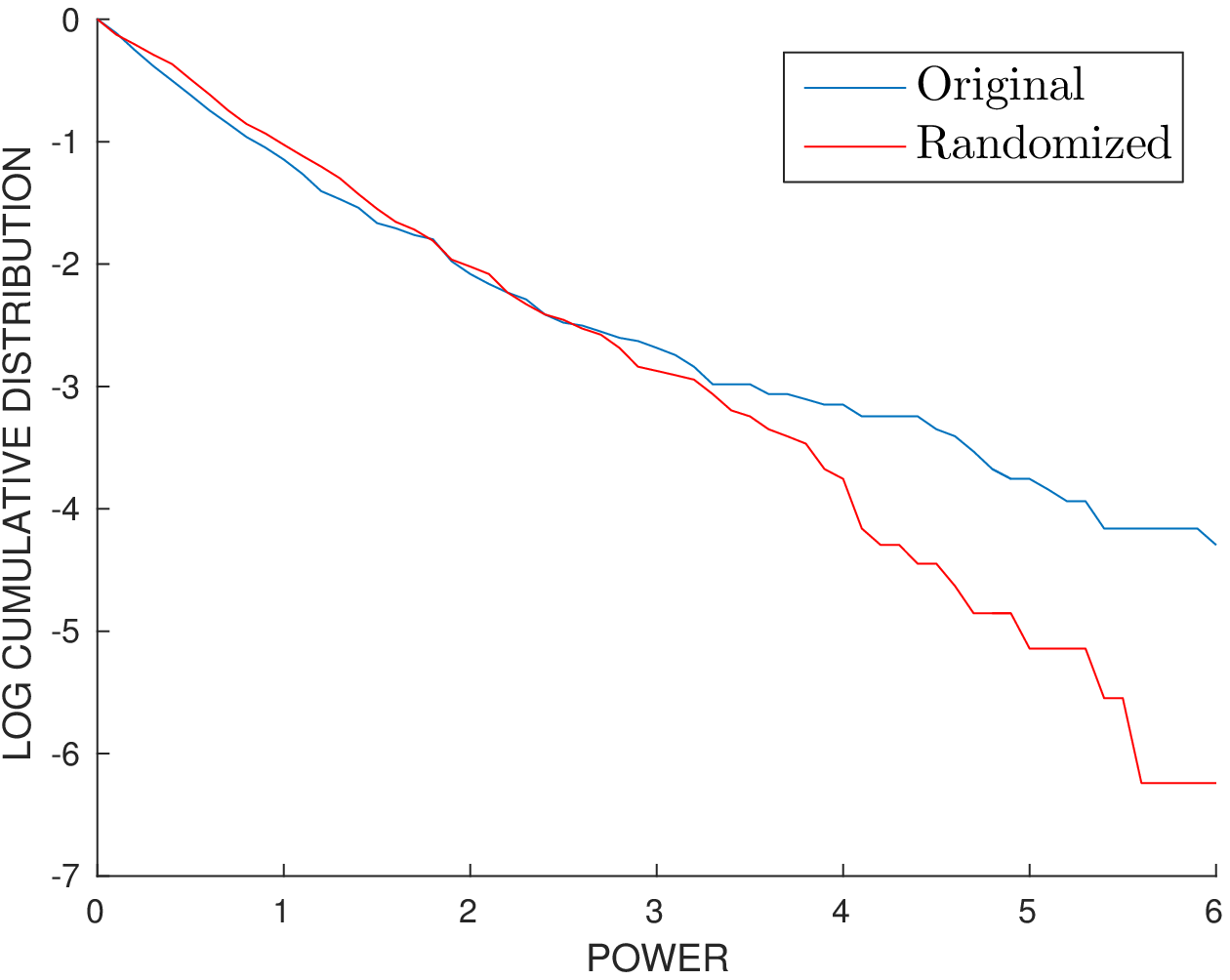}
\includegraphics[width=0.45\textwidth]{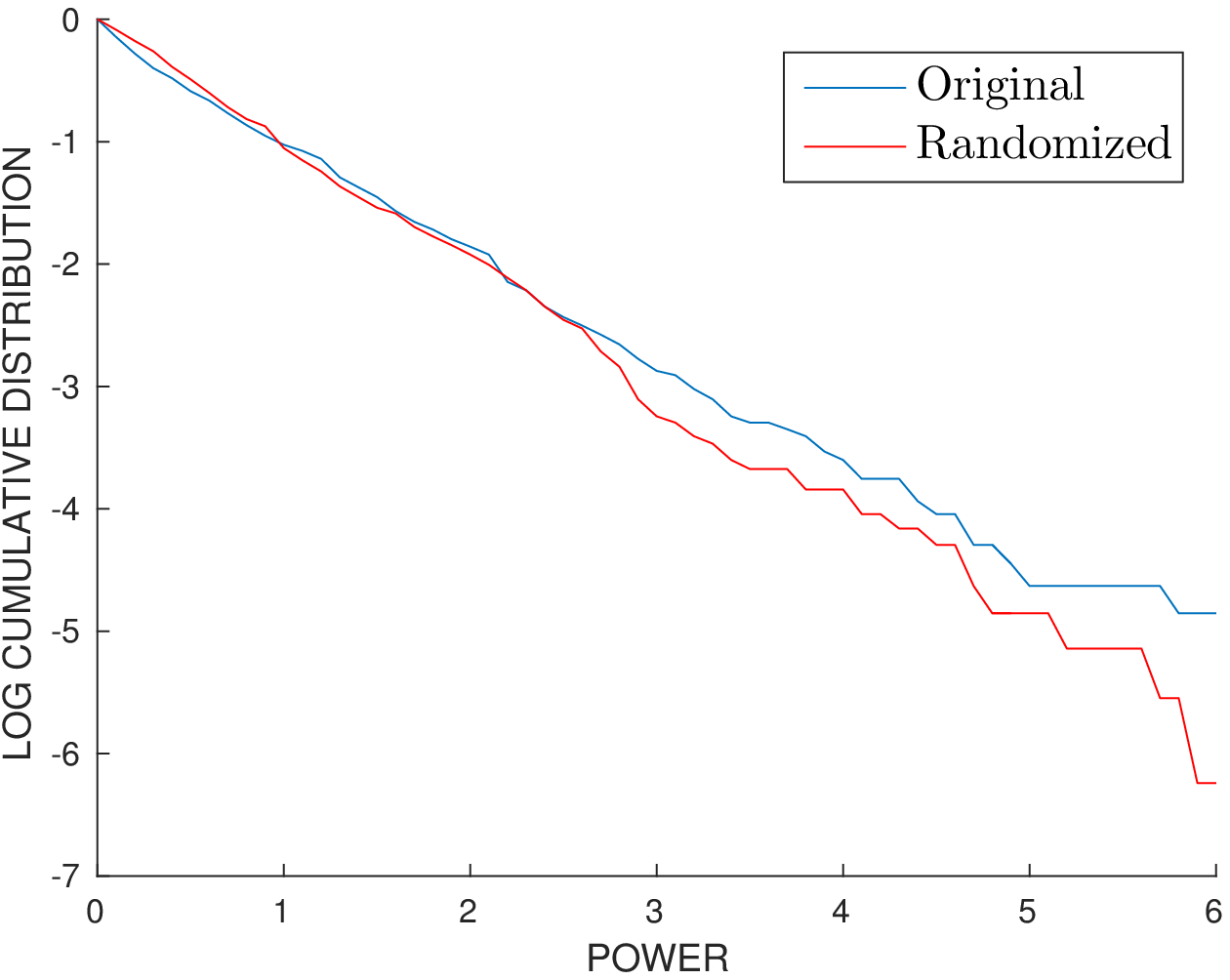}
\caption{Illustration of the method of estimating the significance of Fourier spectral power peaks using randomized datasets for the CME occurrence (left column) and mass (right column) rates. The upper panels display the power spectra with two significance levels. The lower panels display the logarithm of the cumulative probability of the spectra versus power for the original and randomized datasets.
}
\label{fig:test_noise}
\end{figure}

The first approach consists in re-arranging the order of events in the original time series so as to eliminate possible periodicities (and more generally, to eliminate coherent signal) while retaining the noise statistics and in comparing the logarithm of the cumulative distribution of the original and re-arranged series.
In the general case of a normally distributed noise source, its power spectrum has an exponential distribution and thus the logarithm of the cumulative distribution of the power decreases linearly with power.
Departure from this behavior by real data is an indicator of the presence of significant features (\ie coherent periodic signals) and the
slope of the logarithm of the cumulative distribution of power is a measure of the variance $\sigma^2$ of the spectrum \citep{Delache1985}.
For instance, in their application to the detection of solar gravity mode oscillations, \cite{Delache1985} have taken the original data in reversed order.
In their search for CME periodicities, \cite{Lou2003} and \cite{Lara2008} have generated a new time series by randomly re-arranging the order of the events, a process similar to shuffling cards.
In their applications, they found that the original data deviate from random noise for powers larger than 3 to 4$\sigma$ level (see for instance Figure 1 of \cite{Lou2003}), thus providing a threshold for discriminating significant peaks in the spectrum.
Our own implementation of this procedure is displayed in Figure~\ref{fig:test_noise} for both the occurrence and mass rates of CMEs.
Spectrum and confidence levels for 3$\sigma$ and 4$\sigma$ are shown in the upper panels whereas the logarithm of the cumulative distributions of the power spectra of both original and randomized time series  are shown in the lower panels.
In the case of the occurrence rate, a departure of the two distributions is clearly seen at a power corresponding to approximately 3.3$\sigma$ thus allowing to define a significance level threshold.
This is not the case of the mass rate as the two distributions remain close.
Therefore choosing a threshold of 3 or 4$\sigma$ becomes questionable and may lead to retaining non-significant periods.
There are additional problems with this method as it tends to allow a large number of periodicities not supported by other analysis and to produce unwanted peaks exceeding the threshold that may persist in the power spectrum of the randomized time series.

\begin{figure}[htpb!]
\noindent
\centering
\includegraphics[width=\columnwidth]{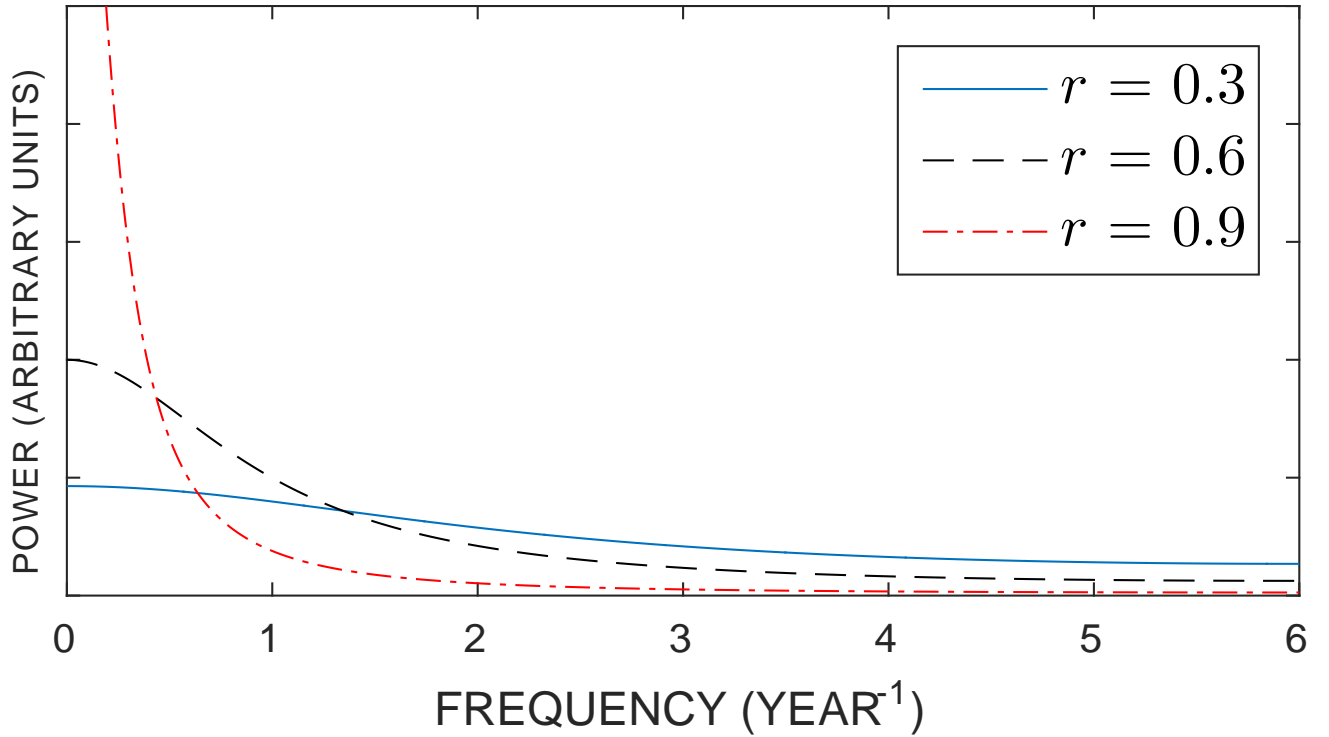}
\caption{Fourier power spectra of red noise models corresponding to correlation coefficients $r=$0.3, 0.6, and 0.9.}
\label{fig:theo_rednoise}
\end{figure}

The second approach consists in choosing a model of the noise among the two classical ones, white or red, most appropriate to the real data.
Past works often state that the red noise model is more appropriate to physical phenomena on the basis of its power spectrum being weighted toward low frequencies, however without formal proof.
Note that this model requires specifying a coefficient of correlation $r$ between two successive time samples so that its ``redness'' depends
upon $r$ which can be adjusted to match the observed time series.
Figure~\ref{fig:theo_rednoise} displays three power spectra of red noise for $r=0.3, 0.6, 0.9$, the last case corresponding to a spectral density inversely proportional to its frequency squared.

\begin{figure}[htpb!]
\noindent
\centering
CME occurrence rate \hspace{3.7cm} CME mass rate\hspace{0.2cm}
\includegraphics[width=0.47\textwidth]{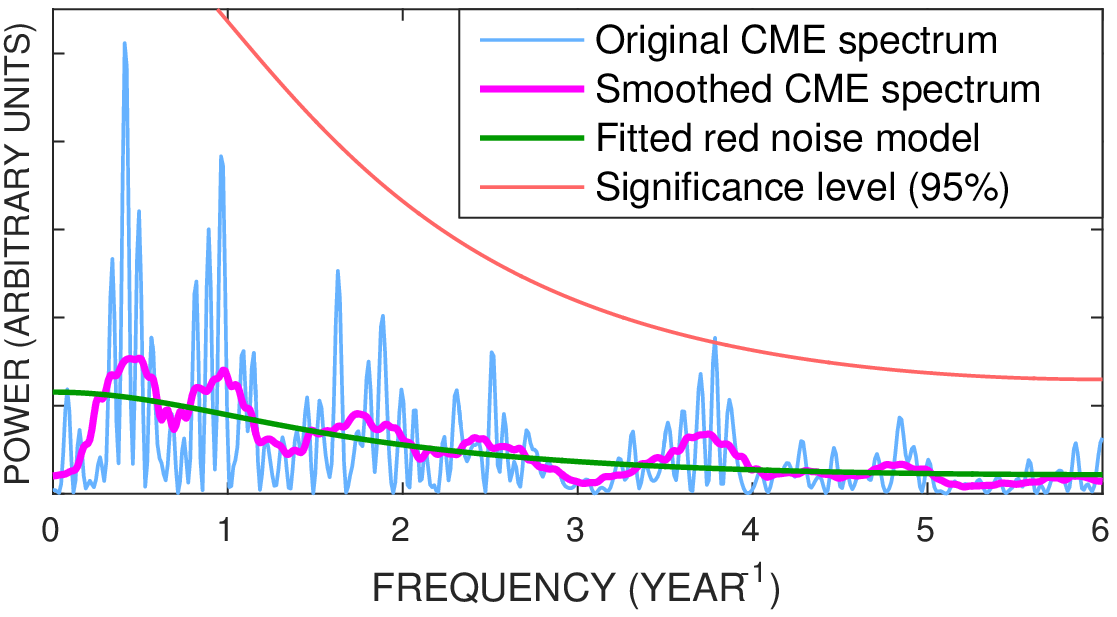} ~~~~
\includegraphics[width=0.47\textwidth]{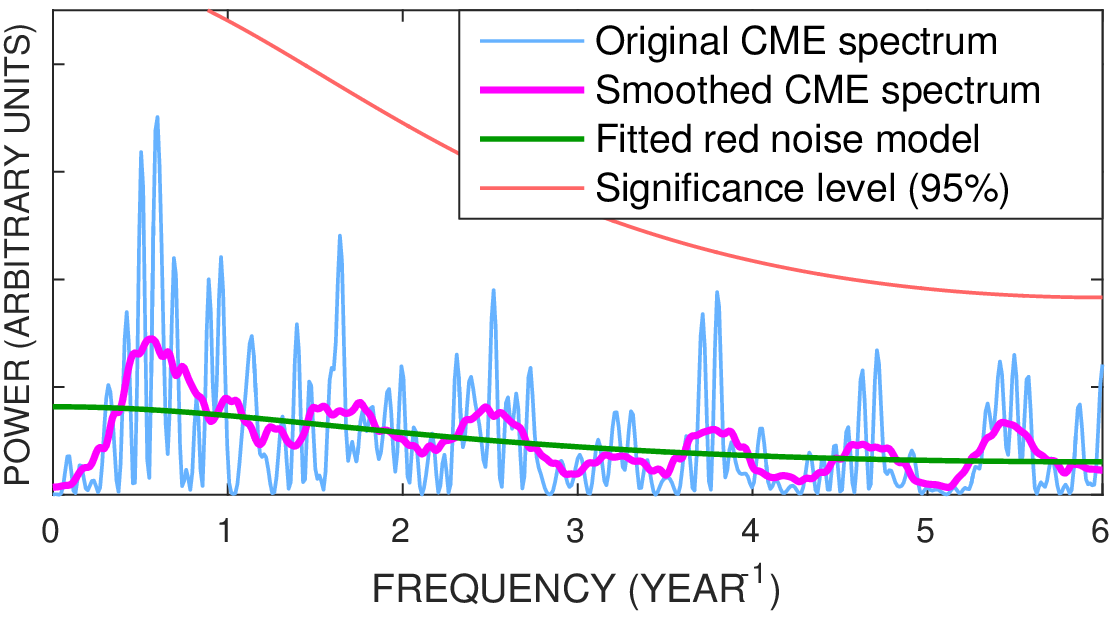}
\caption{Original and smoothed Fourier power spectra of the original datasets and of the fitted red noise models for the CME occurrence (left panel) and mass (right panel) rates.}
\label{fig:red_noise}
\end{figure}

Using the classical Fourier transform allows us to build a statistical test against the white or red noise null hypothesis similar to that developed for wavelet analysis by \cite{Torrence1998} whereas this is not possible with other methods of frequency analysis.
Following these authors, the best fit between the spectrum of the red noise model and that of the observed time series is obtained by taking the lag-1 auto-correlation coefficient of the latter for estimating $r$.
This procedure is applied to the CME occurrence and mass rates and yields $r$=0.35 and $r$=0.30 respectively.
Figure~\ref{fig:red_noise} displays the power spectra of the fitted red noise models (green curves) together with the heavily smoothed power spectra (pink curves) of the CMEs and shows a similar decrease of the spectra with increasing frequency, thus justifying the choice of the red noise model.
\\
It also displays the deduced 95\% significance level (orange curves) computed from the fitted red noise models.
In the following study, we adopt the criterion that a period is statistically significant if the corresponding power exceeds the 95\% significance level against the red noise background as illustrated in Figure~\ref{fig:red_noise}.
Unfortunately, it is not possible to likewise transfer the \cite{Torrence1998} procedure to the other periodogram estimators.
An alternative procedure has been proposed by \cite{Pardo-Iguzquiza2006} in the case of the maximum entropy method.
It is based on a permutation test, which is a distribution-free (but computer intensive) method.
It allows managing both hypotheses of white and red noises but it requires two important parameters: the order of the auto-regressive spectral estimator, and the number of permutations in the test.
They must be chosen empirically, even if there exists some tricks to help the choice which is not obvious \citep{Papoulis1984}.
Finally, it could not be guaranteed that the choice of a particular model of red noise (among the whole family of red noises) is similar to that used in the wavelet analysis.
Based on these difficulties and on our preference of avoiding using different, henceforth inconsistent, methods of generating the red noise models for the frequency and the time-frequency analysis, we decided to present the Welch, window clean, and maximum entropy spectra without significance levels.
However, they remain instructive for comparing with the Fourier spectra and assessing the periodicity peaks.

If the input signal is stationary, the validity of the periodogram is theoretically assured.
But this strong assumption is rarely satisfied in real datasets leading to a ``dilution'' of the power spectra of a frequency present only in a restricted temporal interval of the signal.
To remedy this drawback, one may try to restrict the frequency analysis to time intervals during which the signal is expected to be close to stationary.
Another alternative is to use the time-frequency analysis as described in the next sub-section.

%-------------------------------------------------------------------------------------------------------------------------------------
\subsection{Time-Frequency Analysis: Wavelet Technique}
%-------------------------------------------------------------------------------------------------------------------------------------

Classical Fourier analysis allows the study of a signal only in the frequency domain, whereas wavelet analysis yields information in both time and frequency domains.
That is why this technique has been used in many fields of physics and astrophysics in order to study the temporal behaviour of oscillatory signals.
Continuous wavelet transform (CWT) techniques, as a tool of signal analysis, were developed in the late $1980s$ \citep{Farge1992} and since then, many advances have been made.
Following recent past works, we choose a continuous wavelet transform using the ``Morlet'' mother wavelet (localized wavelet function) and rely on the technique elaborated by \cite{Torrence1998}.
A Morlet mother wavelet is formed by a plane wave modulated by a Gaussian function controlled by the non-dimensional parameter $\omega_0$.
We choose $\omega_0$ = $6.0$ to satisfy the admissibility condition \citep{Farge1992} and because the Fourier period and the wavelet time scale are nearly equal for this value \citep{Jaffard2001}.
In practice, we adapt the wavelet package\footnote{https://github.com/grinsted/wavelet-coherence} developed by A.~Grinsted to our application and adjust the parameters to obtain a satisfactory compromise between frequency and temporal resolutions.
The result of the wavelet analysis of a given dataset is visualized as a time-frequency spectrum.
It suffers from edge effects at both ends of the time series, restricting the domain of validity whose duration depends upon the period.
These restrictions define the so-called ``cone of influence'' which limits the time-frequency domain.
As also introduced in many past works, we generate global wavelet spectra by time-averaging the power in each frequency step limited by the cone of influence.
The time information of wavelet analysis is lost in this summing process, so that global wavelet spectra are directly comparable to the periodograms (henceforth, they will be displayed altogether).
To some extent, global wavelet spectra may be viewed as a bridge between the results of pure frequency analysis and those of time-frequency analysis.
Consistent with the frequency analysis, the criterion of statistically significant signals at the 95\% level against the red noise background is introduced in both time-frequency spectra (it defines closed sub-domains in these spectra) and global wavelet spectra following the technique of \cite{Torrence1998}.

%=====================================================================================================================================
\section{Periodicities in the CME Occurrence and Mass Rates} \label{Per_CME}
%=====================================================================================================================================

Figure~\ref{fig:plots_CME} displays the temporal evolution of the monthly occurrence rates of the different groups CME, CME$_\textrm{m}$, CME$_\textrm{N}$
and CME$_\textrm{S}$.
It clearly reveals that, in addition to following the $\approx$11-year Schwabe solar cycle, they exhibit oscillations at higher frequencies prominently present during the maxima of SC 23 and 24 but hardly visible during the minima since the rates drop to very small numbers.
We now search for periodicities in:
i) the occurrence rates of the global set of CMEs and of the two subgroups CME$_\textrm{N}$ and CME$_\textrm{S}$, and
ii) the mass rates of the three subgroups CME$_\textrm{m}$, CME$_\textrm{m,N}$ and CME$_\textrm{m,S}$.

\begin{figure}[htpb!]
%\vspace{0.75cm}
\noindent
\centering
\includegraphics[width=0.49\textwidth,height=5.6cm]{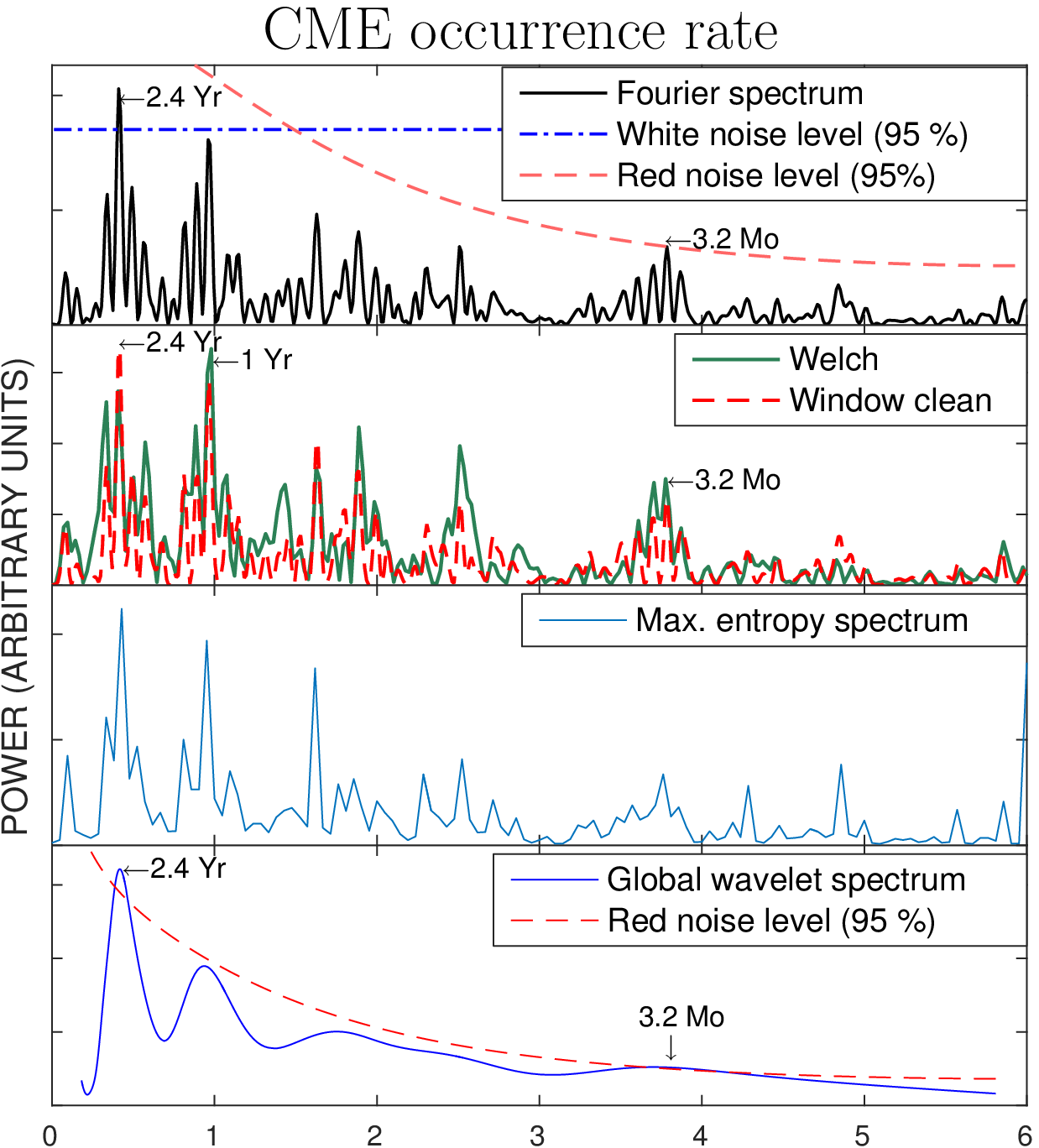}
~\includegraphics[width=0.49\textwidth,height=5.6cm]{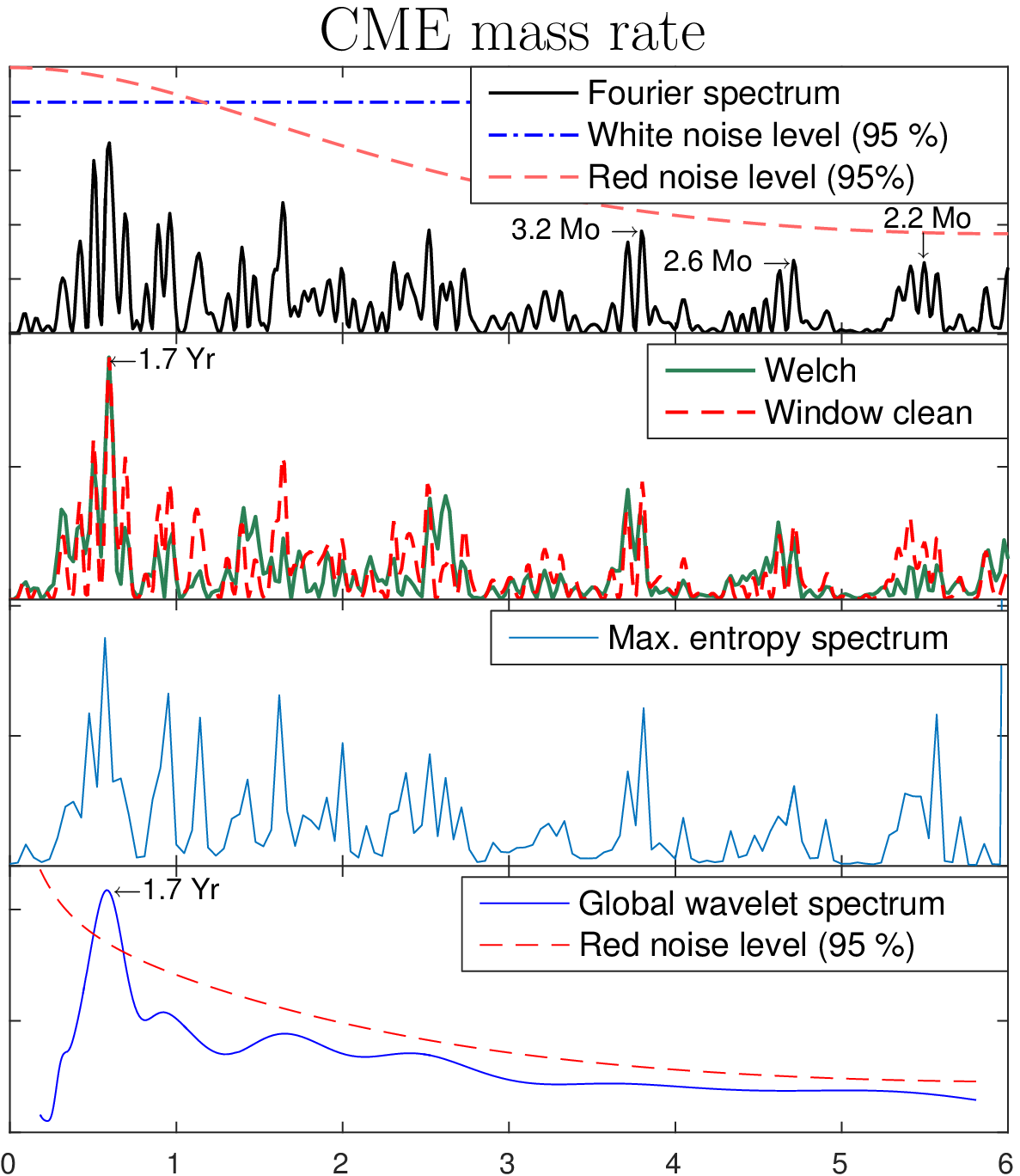} \\
\vspace{0.1cm}
\includegraphics[width=0.49\textwidth,height=5.2cm]{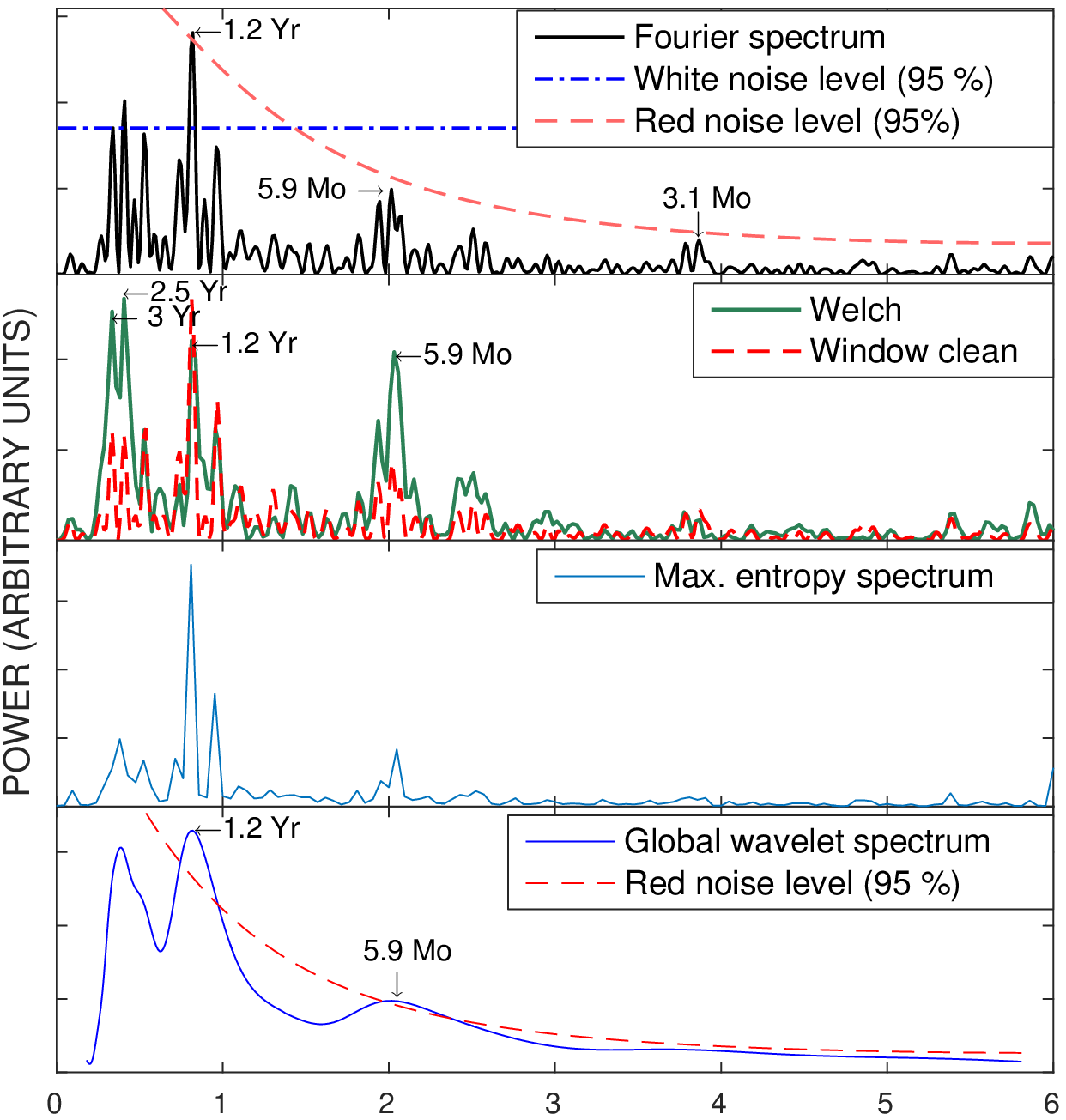}
~\includegraphics[width=0.49\textwidth,height=5.2cm]{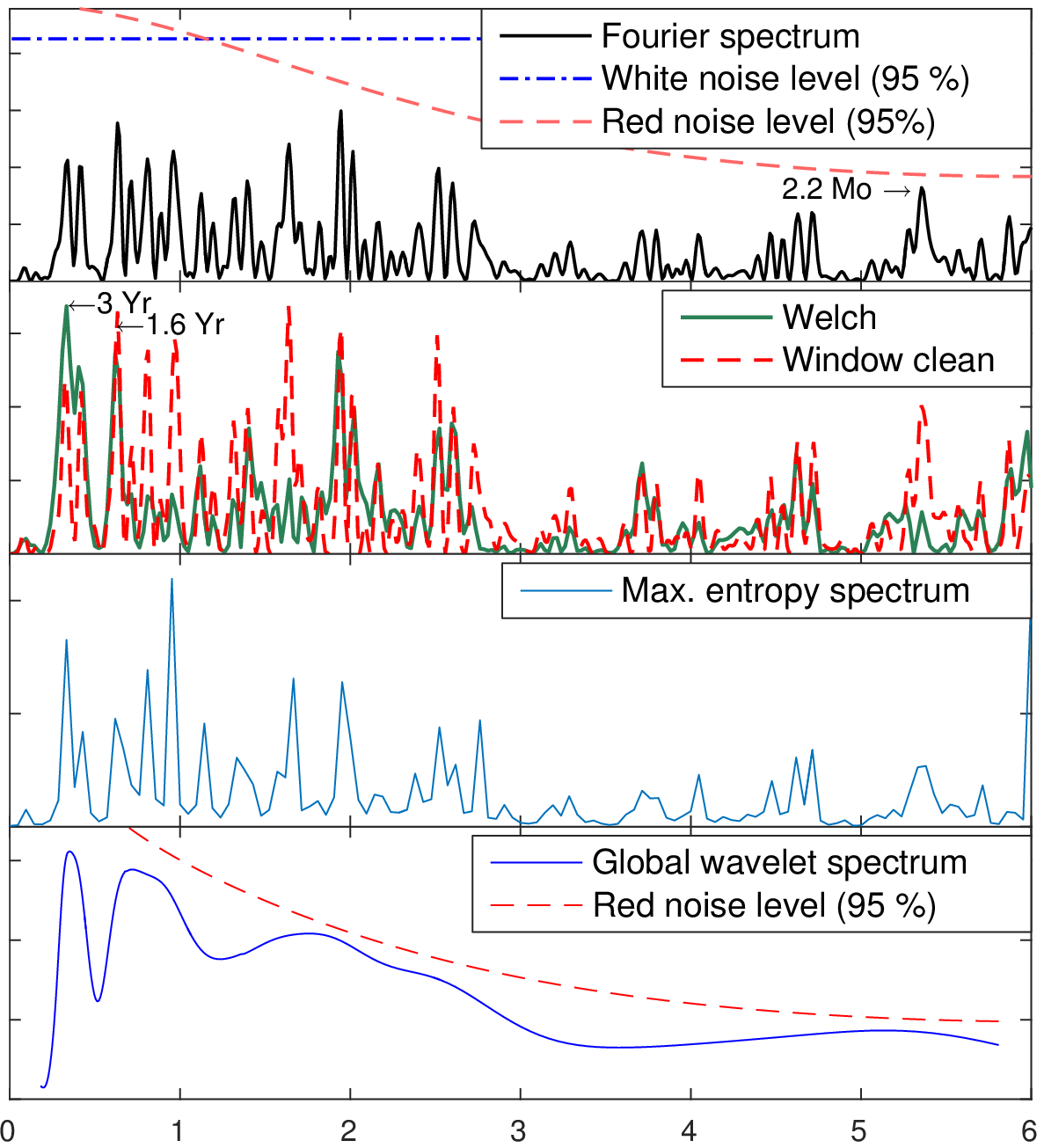}\\
\vspace{0.1cm}
\includegraphics[width=0.49\textwidth,height=5.2cm]{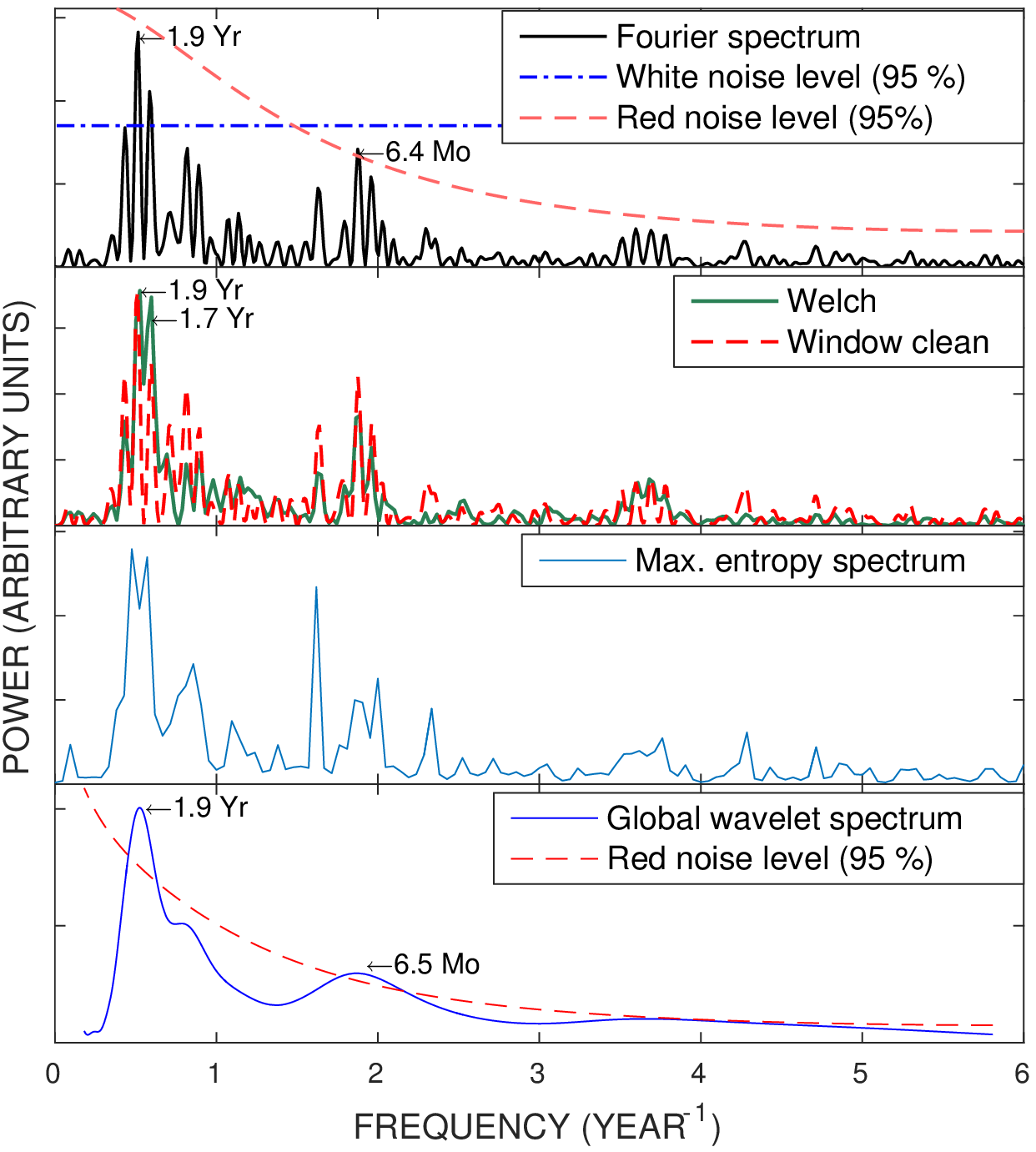}
~\includegraphics[width=0.49\textwidth,height=5.2cm]{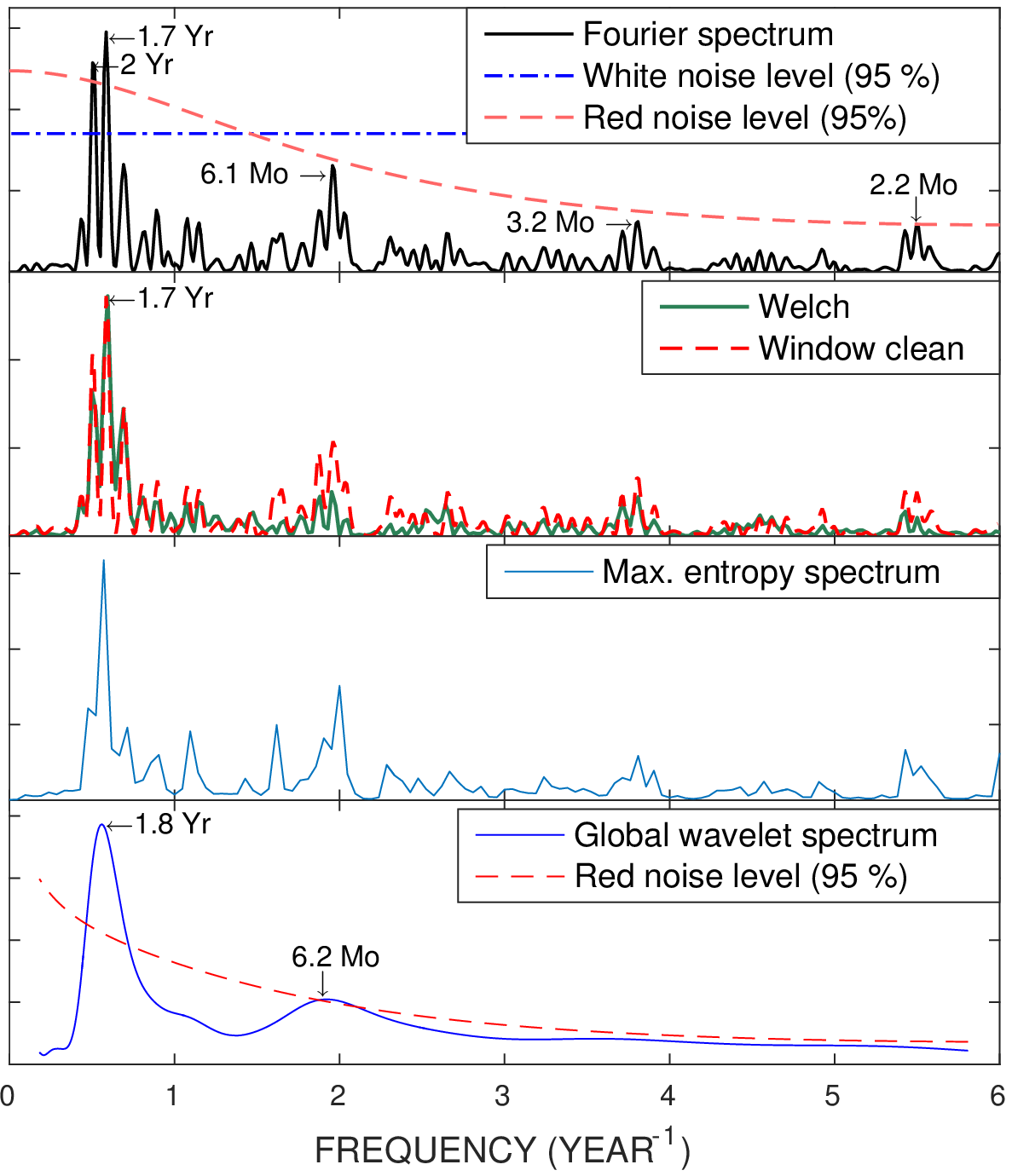}
\caption{Periodograms and global wavelet spectra of monthly CME occurrence  and mass  rates globally (upper panels), in the northern (middle panels) and southern (lower panels) hemispheres.
The most significant peaks are labeled in either month (Mo) or year (Yr) as most appropriate.
The 95\% significance levels against the red and white noise backgrounds are shown by dashed red curves and dash-dot blue lines respectively.}
\label{fig:periodograms_CME}
\end{figure}

The results of the four selected methods of frequency analysis (Schuster, Welch, the maximum entropy method, and the window clean estimation) are displayed in Figure~\ref{fig:periodograms_CME}.
It is worth emphasizing the general consistency between the results obtained with the different methods.
We note a couple of minor discrepancies affecting the maximum entropy method, namely the low peaks at 2.5 and 3 years in the case of the occurrence rate in the northern hemisphere and the mere absence of a peak at 2 years in the case of the mass rate in the southern hemisphere.
Otherwise, there is an excellent agreement on the peak periods between the four methods.
The shortest observed period of 2.2 months is only seen in the CME$_\textrm{m}$ and its hemispheric subgroups (CME$_\textrm{m,N}$ and CME$_\textrm{m,S}$) at a level slightly below the red noise significance level.
Next, we find a period of 3.2 months which reaches the red noise significance level for the occurrence rate of the global set of CMEs, but is absent in all other cases.
It can however been seen slightly below the significance level in the CME$_\textrm{N}$ and CME$_\textrm{m,S}$ subgroups.
A period 6.4 months is present in the southern hemisphere CMEs, but is well below the significance level in all other cases; however, close periods of 5.9 and 6.1 months barely miss the red noise criterion for respectively the CME$_\textrm{N}$ and CME$_\textrm{m,S}$ subgroups.

A set of mid-range oscillations is observed ranging from 1 to 2 years.
Periods of 1.2, 1.7, and 2 years meeting the red noise criterion are detected in the CME$_\textrm{N}$ subgroup for the first one and in the CME$_\textrm{m,S}$ subgroup for the other two.
An additional peak close to meeting the red noise criterion of 1.9 year can be further discerned in the CME$_\textrm{S}$ subgroup.

\begin{figure}[htpb!]
\noindent
\centering
\includegraphics[width=\textwidth,height=0.81\textheight]{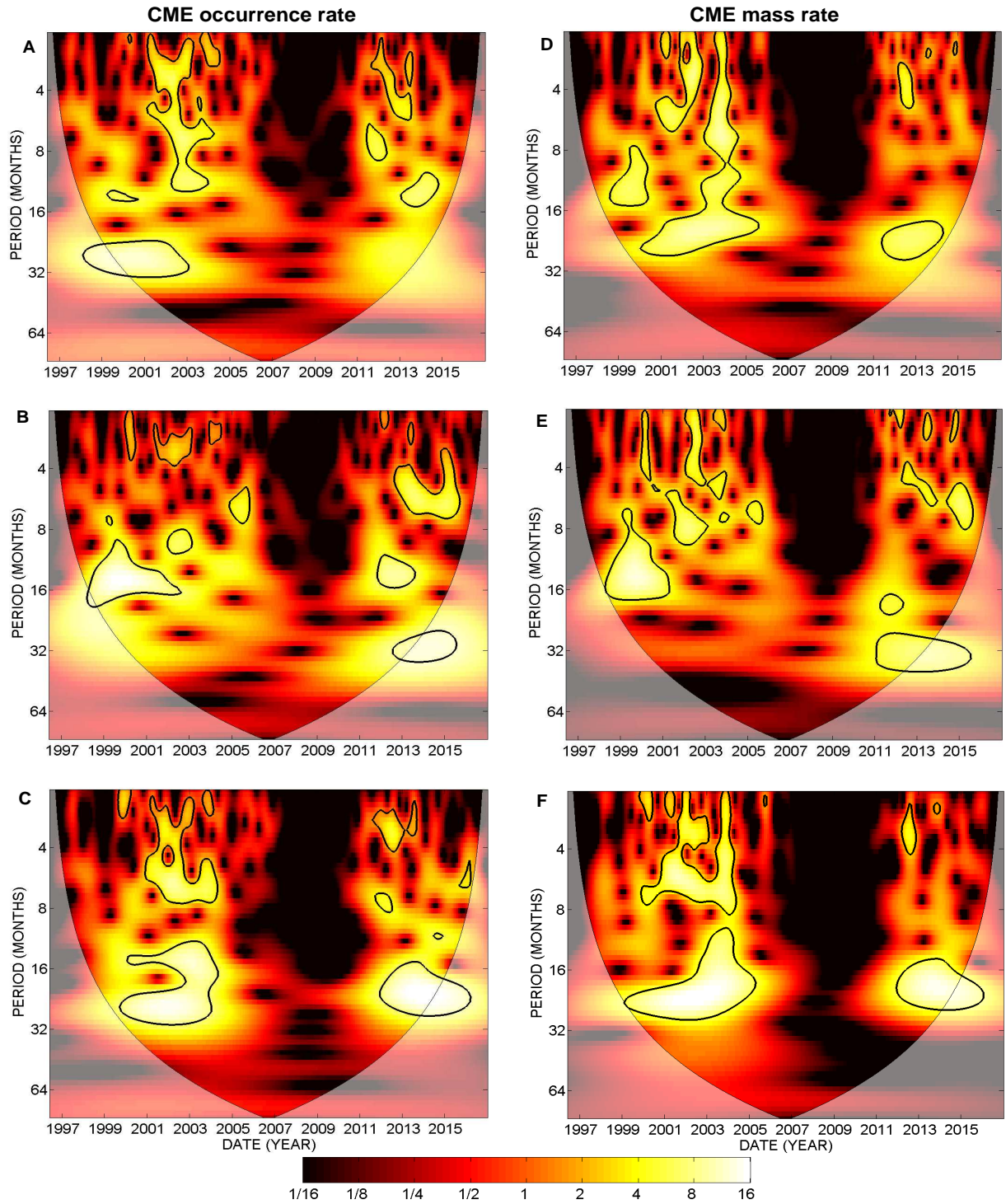}
\caption{Wavelet spectra of monthly CME occurrence (left column) and mass (right column) rates globally (upper panels), in the northern (middle panels) and southern (lower panels) hemispheres.
The shaded regions affected by edge effects delimit the cones of influence.
Regions where the spectral power is statistically significant at the 95\% level against the red noise backgrounds are contoured by black, thick lines.
The color bar provides the scale for the wavelet power amplitude (arbitrary units).}
\label{fig:wt_CME}
\end{figure}

Figure~\ref{fig:wt_CME} displays six wavelet spectra
-- CME occurrence and mass rates and their associated north and south subgroups --
resulting from the wavelet spectral analysis.
Here and in the following wavelet spectra, wavelet power amplitudes are quantified by color bars where the color ranges from black (minimum power) to bright yellow (maximum power), and zones affected by edge effects (cones of influence) are shaded.
Consistent with the criterion adopted in the frequency analysis, regions where the spectral power is statistically significant at the 95\% level against the red noise backgrounds are contoured by black, thick lines.
These results confirm those of the frequency analysis, but show more precisely the time intervals during which the various periods prevail.
The most stable and statistically significant maxima of the wavelet power amplitude broadly extend over one to two years (depending upon the case but consistent with the frequency analysis, for instance the oscillations at 1.2, 1.7, 1.9 and 2 years already identified in the periodograms) and prominently during the maximum of SC 23, and less so during that of SC 24.
Higher frequency oscillations (with periods less than one year) appear as statistically significant ``islands'' and here again prominently during the maximum of SC 23, and less so during that of SC 24.

The global wavelet spectra are displayed in Figure~\ref{fig:periodograms_CME} to allow a direct comparison with the spectra produced by the frequency analysis.
As expected, they represent highly smoothed versions of the latter spectra.
Indeed, the averaging process results in a merging of the close peaks in broad enhanced maxima that in some cases exceed the red noise criterion whereas the individual peaks did not.

\begin{figure}[htpb!]
\noindent
\centering
\includegraphics[width=0.45\textwidth]{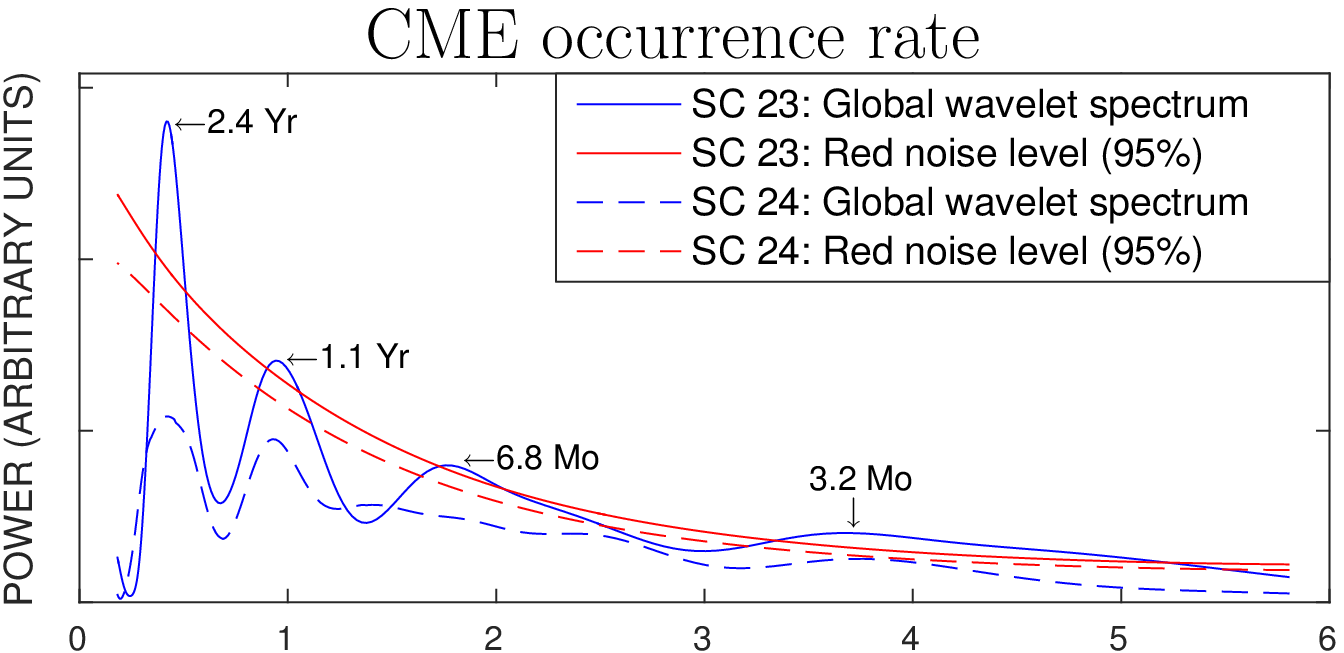}
~~~~~~\includegraphics[width=0.44\textwidth]{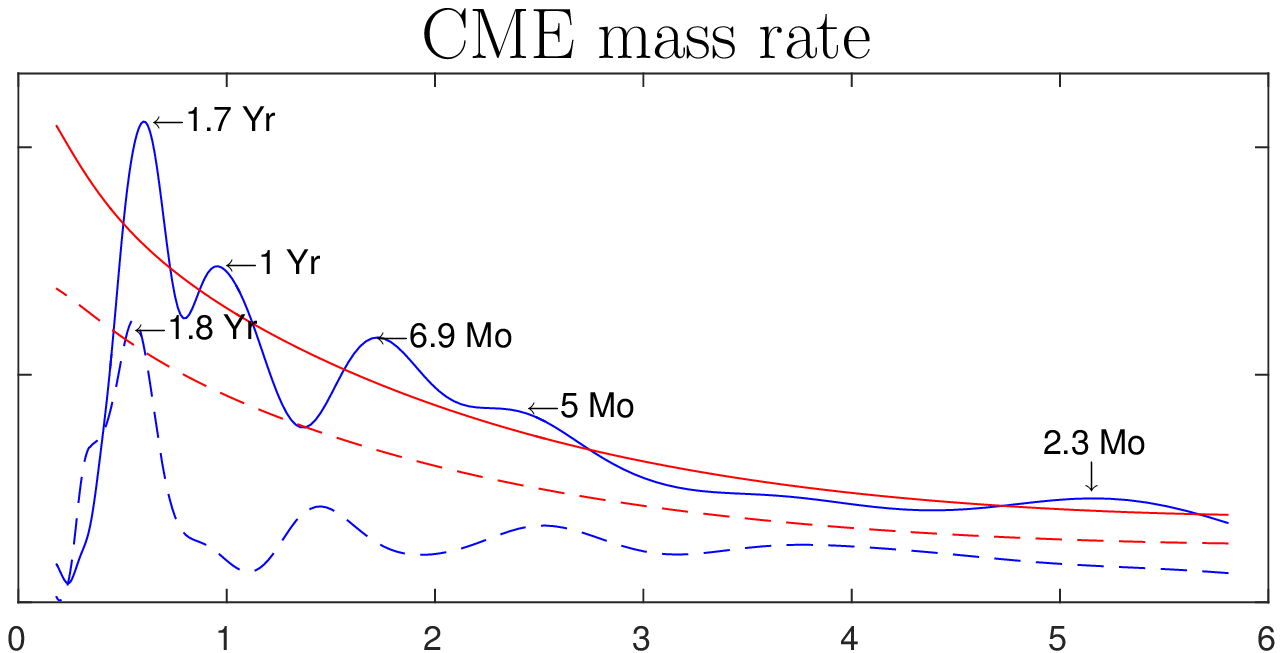}\\
~\\
\includegraphics[width=0.45\textwidth]{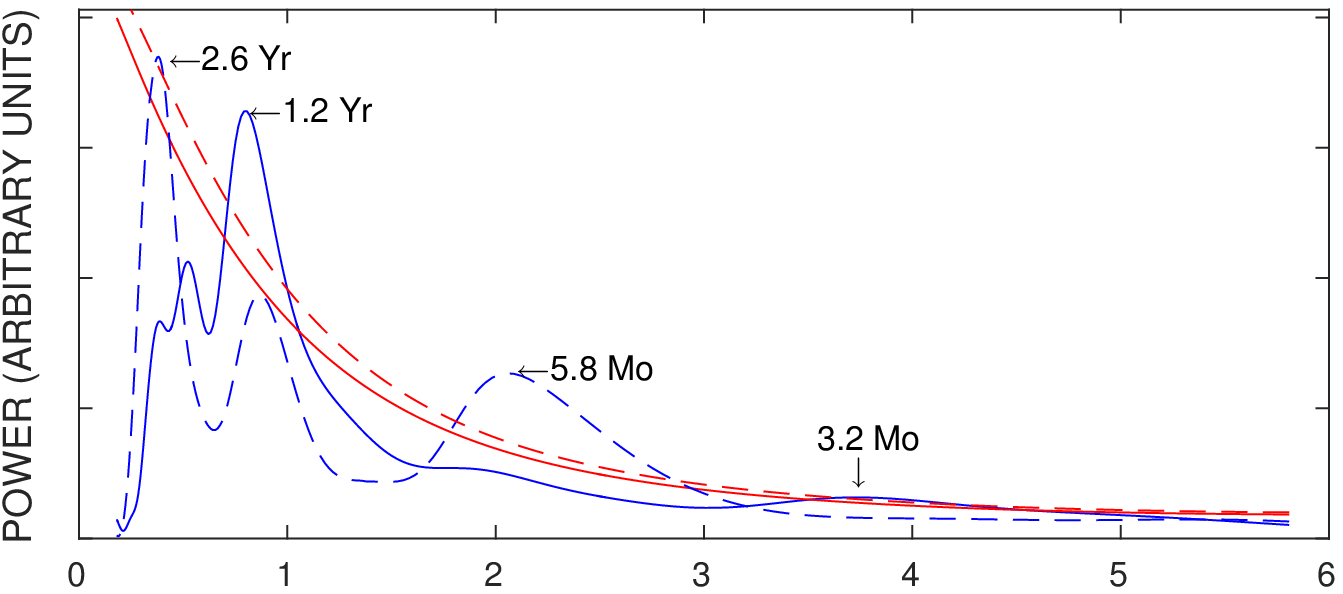}
~~~~~~\includegraphics[width=0.44\textwidth]{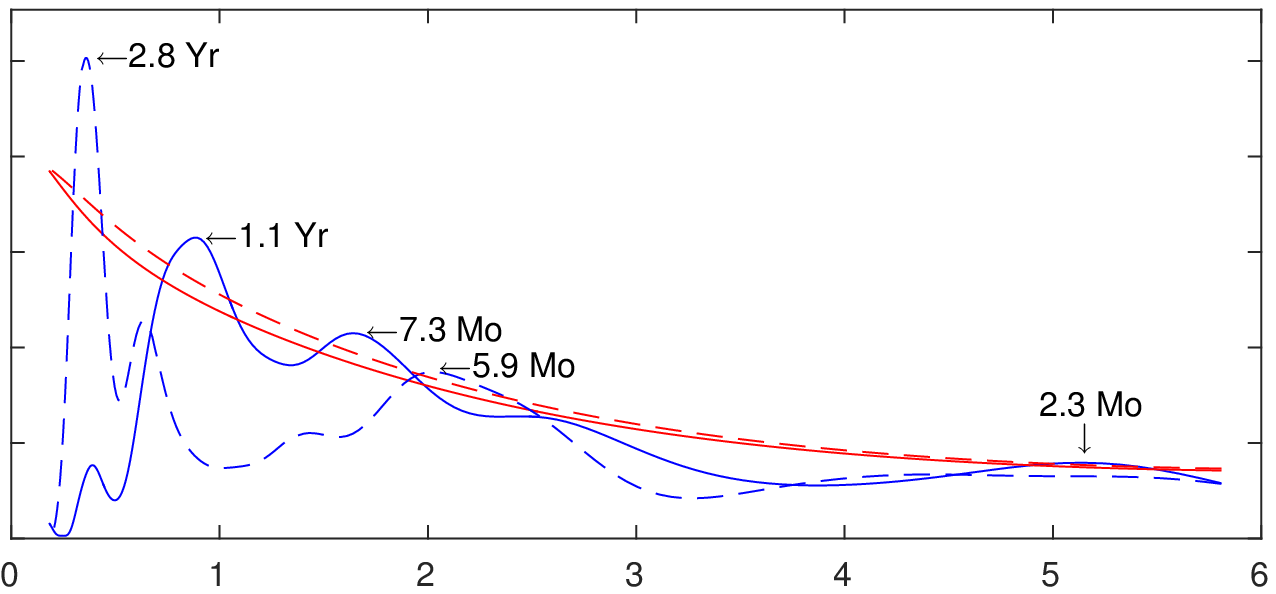}\\
~\\
\includegraphics[width=0.45\textwidth]{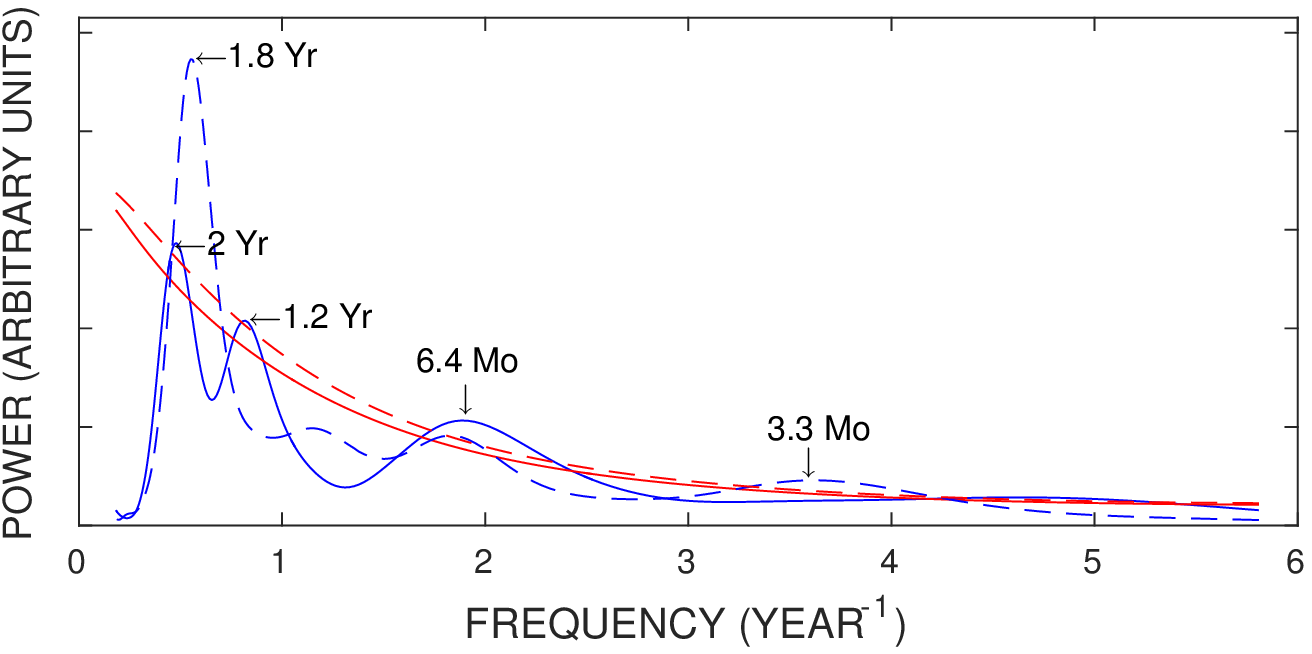}
~~~~~~\includegraphics[width=0.44\textwidth]{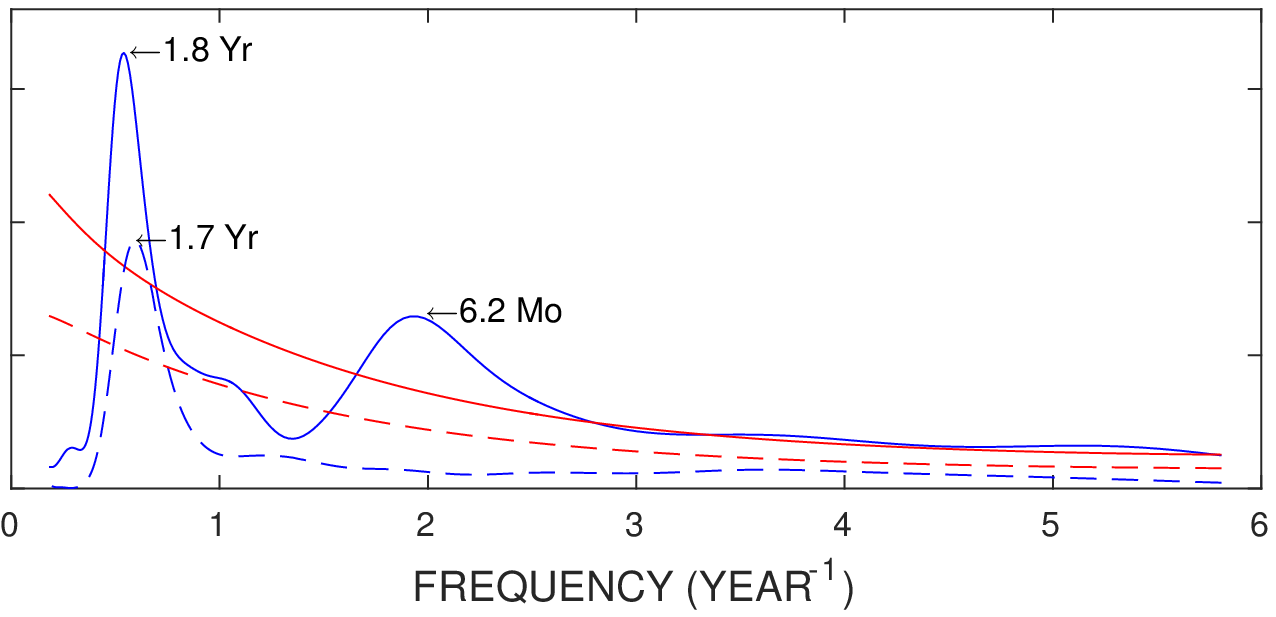}
\caption{Global wavelet spectra of monthly CME occurrence (left column) and mass (right column) rates for Solar Cycles 23 and 24 globally (upper panels), in the northern (middle panels) and southern (lower panels) hemispheres.
The spectra are shown by blue curves and the 95\% significance levels against the red noise backgrounds by red curves.
The continuous and dashed lines correspond to SC 23 and 24 respectively.}
\label{fig:gwt_CME_by_cycle}
\end{figure}

In Figure~\ref{fig:gwt_CME_by_cycle}, we split the global wavelet spectra in two time intervals corresponding to the maximum phase of SC 23 and 24 thus allowing to refine our analysis.
It can be seen that, in the case of the occurrence rate of the global set of CMEs, the period of 3.2 months is confirmed for the two maxima.
There however appears periods of 6.8 months, 1.1 and 2.4 years in SC 23 which are absent in SC 24.
In the case of the CME$_\textrm{N}$ subgroup, periods of 3.2 months and of 1.2 year are present in SC 23 whereas a broad peak centered at 5.8 months and a period of 2.6 years are present in SC 24.
In the case of the CME$_\textrm{S}$ subgroup, the 6.4-month period is observed in both SC 23 and SC 24. SC 23 exhibits additional periods of 1.2 and 2 years whereas SC 24 has 3.3 months and 1.8 year.
Finally, turning our attention to the mass rate of the global set of CMEs, we find a wealth of periodicities in SC 23: 2.3, $\approx$5, and 6.9 months and 1 and 1.7 year, contrary to SC 24 where only one periodicity of 1.8 year is observed.
In the case of the CME$_\textrm{m,N}$ subgroup, several periods are significant: a broad peak centered at $\approx$2.3 months, 7.3-month and 1.1-year periods in SC 23
and 5.9-month and 2.8-year periods in SC 24.
In the case of the CME$_\textrm{m,S}$ subgroup, the 6.2-month and 1.8-year periods are only present in SC 23 whereas a single nearby period of 1.7 year is observed in SC 24.

%===================================================================================================
\section{Periodicities in Proxies of Solar Activity, Flares and Prominences} \label{Per_Others}
%===================================================================================================

Having ascertained the periodicities in the CME activity, we now turn our attention to the selected proxies of solar activity and the two erupting processes.
In order to avoid any bias that could distort the comparison, we apply the same period searching method as used for the CMEs limited to the discrete Fourier transform (as implemented by \cite{Schuster1898}) for conciseness since it allows establishing the significance level against the red noise background and to the time-frequency analysis.
We emphasize that the red noise model is fitted independently to each time series as described in Section \ref{Meth}.

%---------------------------------------------------------------------------------------------------
\subsection{Periodicities in Selected Solar Proxies} \label{Per_Proxies}
%---------------------------------------------------------------------------------------------------

\begin{figure}[htpb!]
\vspace{0.5cm}
\noindent
\centering
\includegraphics[height=\textwidth, width=0.85\textheight,angle=90]{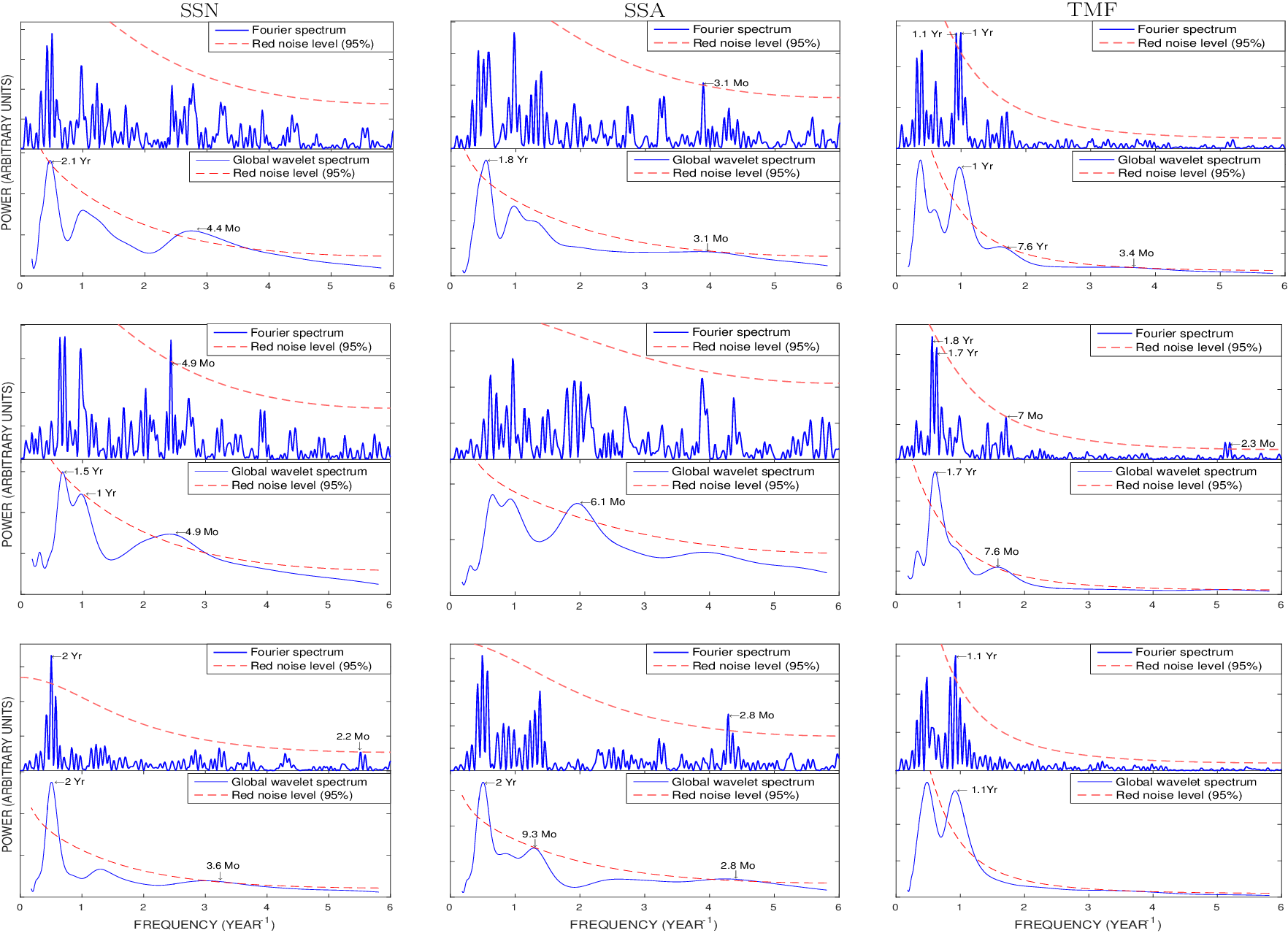}
\caption{Periodograms and global wavelet spectra of SSN (left column), SSA (middle column), and TMF (right column) globally (upper row), in the northern (middle row) and southern (lower row) hemispheres.
The 95\% significance levels against the red noise backgrounds are shown by dashed red curves.}
\label{fig:periodograms_Proxies}
\end{figure}

The results of the frequency analysis of the selected solar proxies are displayed in Figure~\ref{fig:periodograms_Proxies}.
The spectra reveal a ``forest'' of periods without any clear consistent pattern among them.
Those which meet the red noise criterion are: 2.2 months for SSN$_{south}$, 2.3 months for TMF$_{north}$, 2.8 months for SSA$_{south}$, 3.1 months for SSA, 4.9 months for SSN$_{north}$, 7 months for TMF$_{north}$, 1 year for TMF, 1.1 year for TMF$_{south}$, and 2 years for SSN$_{south}$.

\begin{figure}[htpb!]
  \noindent\centering
\rotatebox{90}{
\includegraphics[height=\textwidth, width=0.91\textheight]{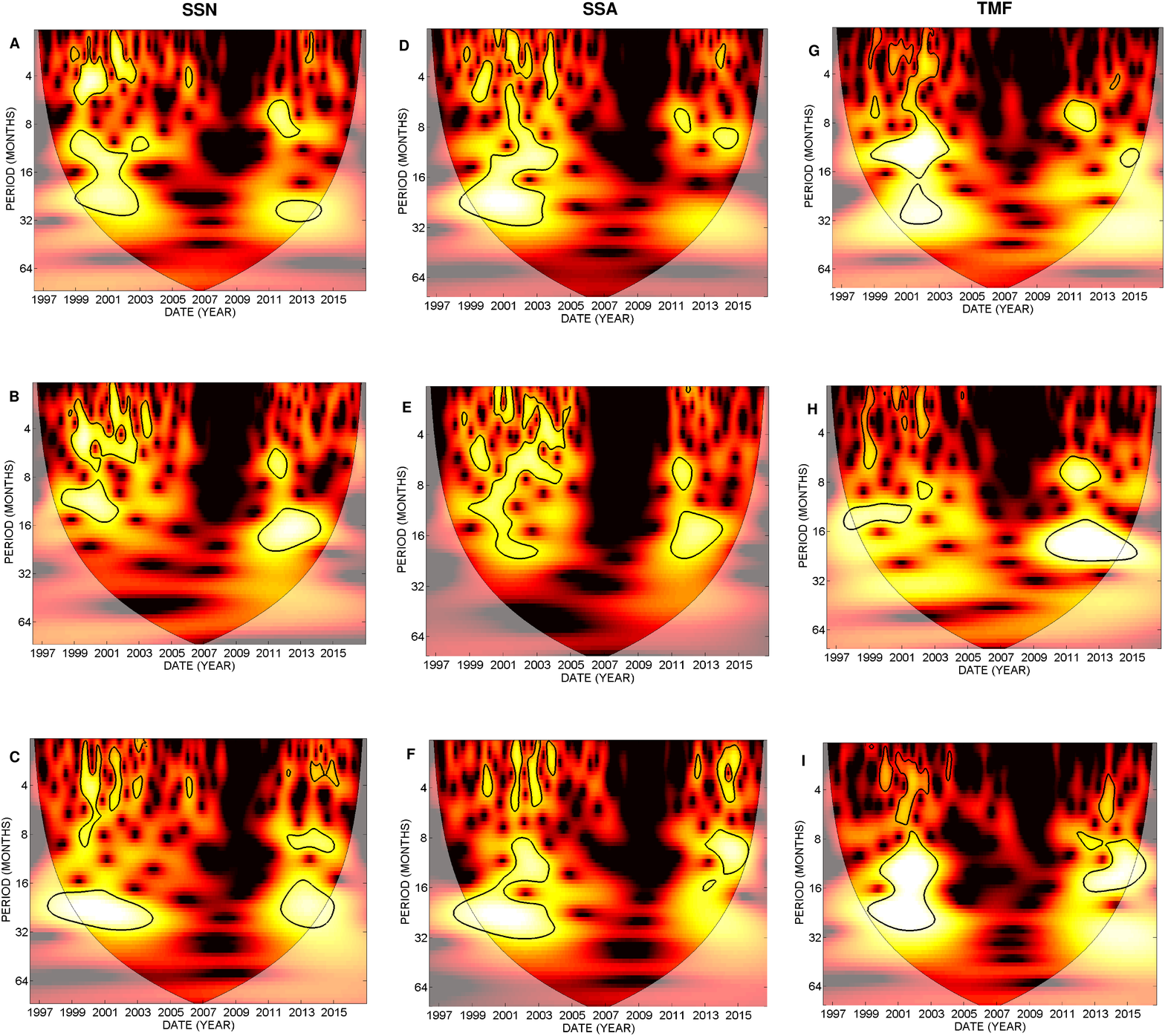}}
\caption{Wavelet spectra of SSN (left column), SSA (middle column), and TMF (right column) globally (upper row), in the northern (middle row) and southern (lower row) hemispheres.
See top of Figure~\ref{fig:wt_CME} for the color bar.}
\label{fig:wt_Proxies}
\end{figure}
Very much like the case of the CMEs, the wavelet spectra of the selected solar proxies globally, and in the northern and southern hemispheres (Figure~\ref{fig:wt_Proxies}) confirm the results of the periodograms and highlight the time intervals where significant periods are present, prominently during SC 23 and less so during SC 24.
The most stable and statistically significant maxima of the wavelet power amplitude are seen at periods in the range 1 to 2 years, the 2-year period being most conspicuous in the southern hemisphere, whereas periods of 2 to 3 months appear as ``islands'' prominently during the maximum of SC 23.
The resulting global wavelet spectra displayed in Figure~\ref{fig:periodograms_Proxies} indicate that a few additional periods meet the red noise criterion:
4.4 months and 2.1 years for SSN, 1 and 1.5 year for SSN$_{north}$, 2 years for SSN$_{south}$, 1.8 year for SSA, 6.1 months for SSA$_{north}$, 9.3 months and 2 years for SSA$_{south}$, and 7.6 months for both TMF and TMF$_{north}$.
We note that the 154-day ($\approx$5.1 months) ``Rieger'' periodicity \citep{Rieger1984} is only marginally found in only one case, namely 4.9 months for SSN$_{north}$.
This ``Rieger'' periodicity was observed in solar flare occurrence during SC 19 to 21 but was absent in SC 22 and 23 \citep{Bai2003}.
The same conclusion has been reached for sunspot areas by \cite{Oliver1998}.
This absence is totally consistent with the mere lack of this periodicity in our data.

%---------------------------------------------------------------------------------------------------
\subsection{Periodicities in Solar Flares} \label{Per_Flares}
%---------------------------------------------------------------------------------------------------

\begin{figure}[htpb!]
\vspace{0.35cm}
\noindent
\centering
\includegraphics[width=0.42\textwidth]{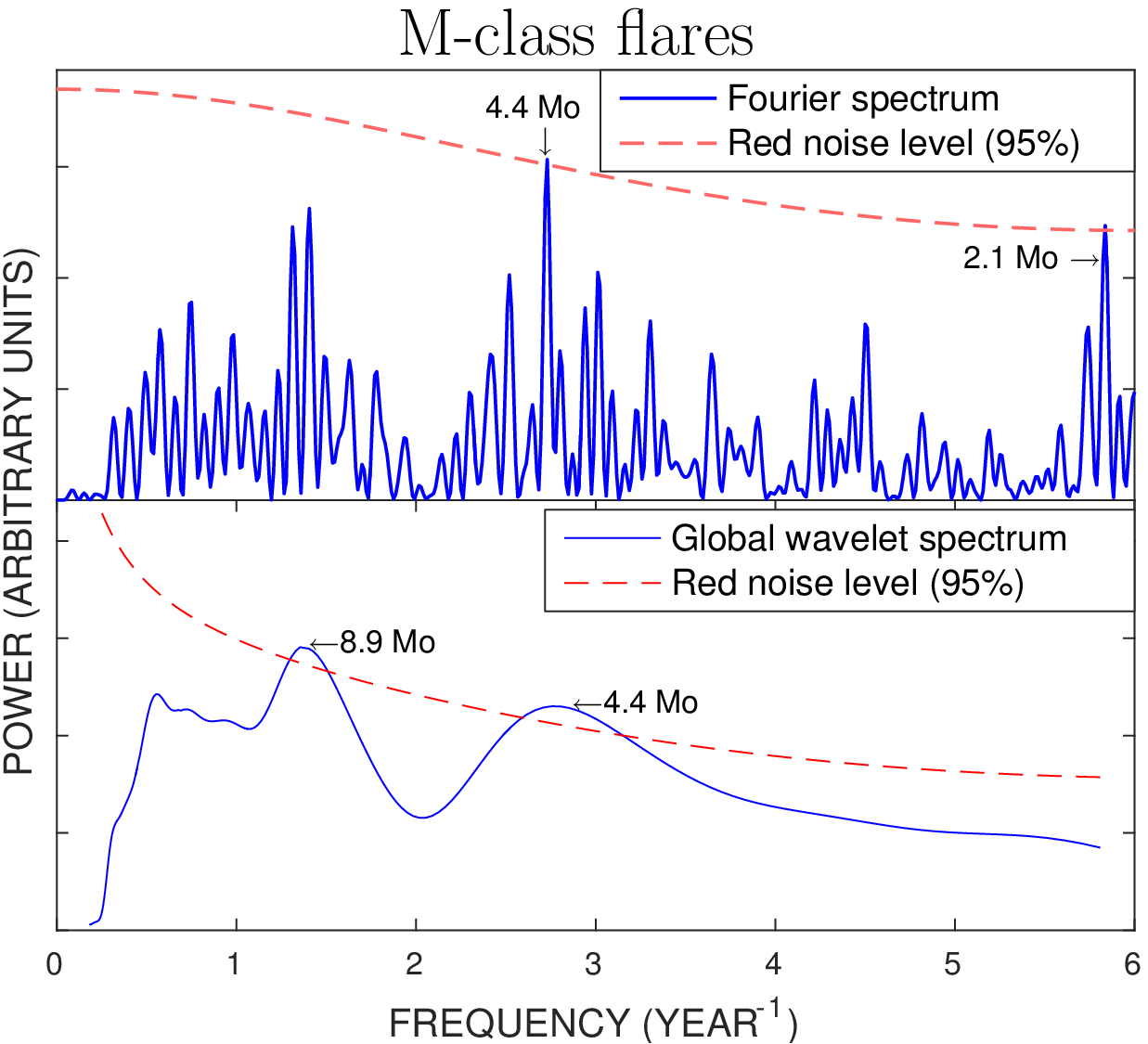}
~~~~~\includegraphics[width=0.41\textwidth]{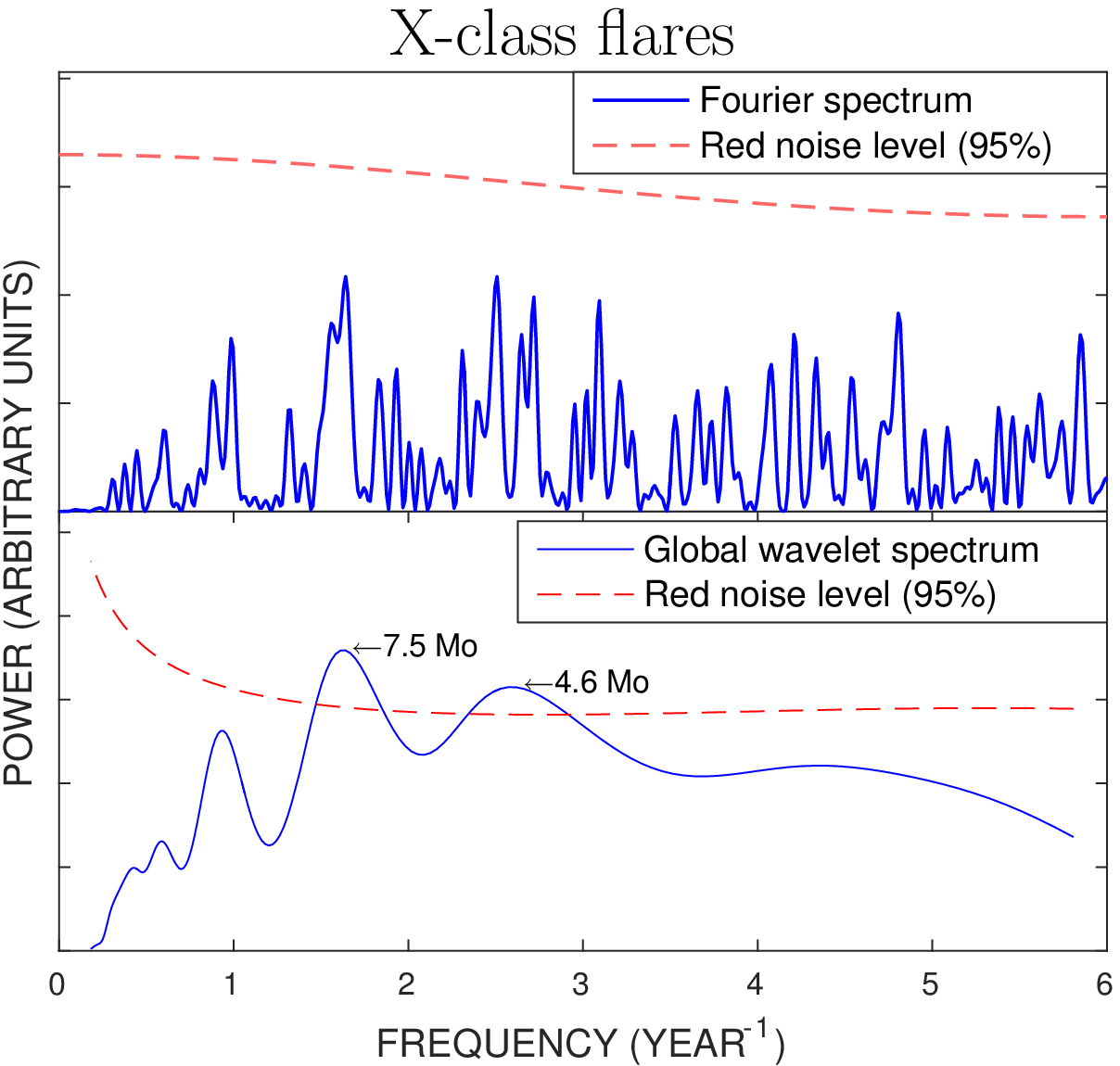}
\includegraphics[width=\textwidth]{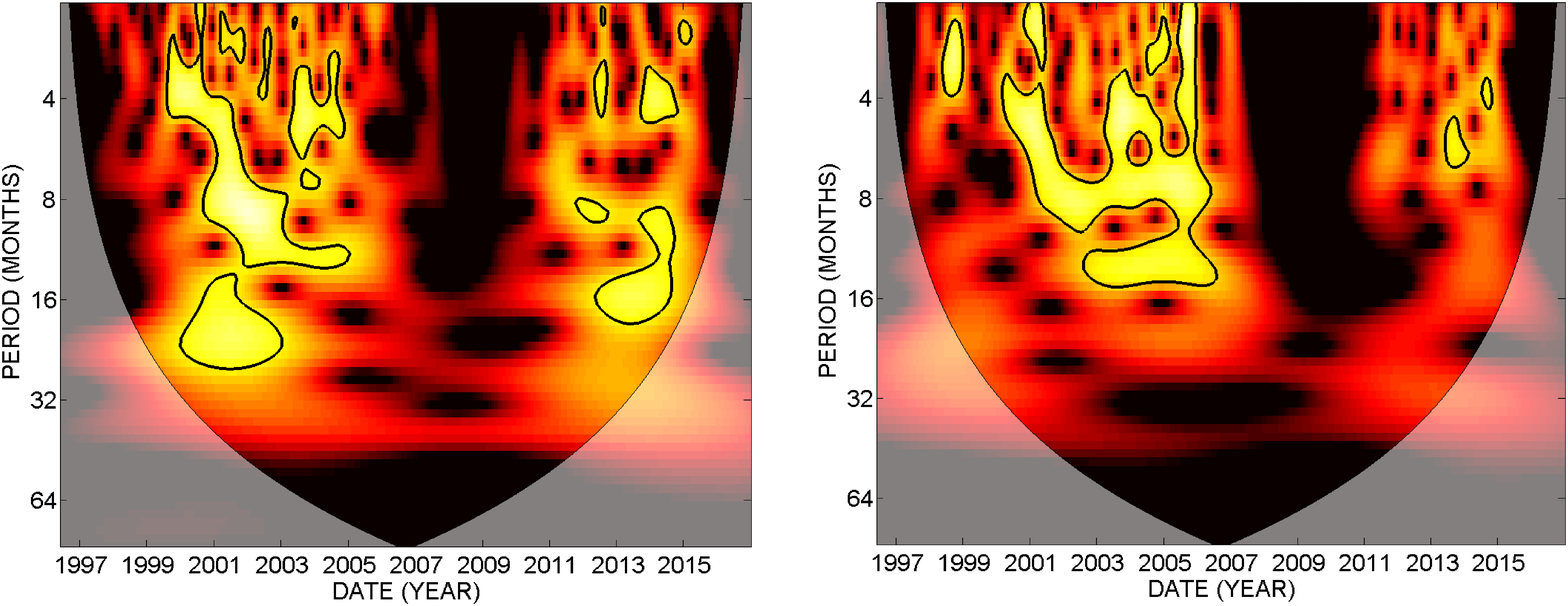}
\caption{Periodicities in the monthly occurrence rate of M-class (left column) and X-class (right column) flares.
Each upper panel regroups two superimposed graphs corresponding to the periodograms (top) and to the global wavelet spectra (bottom).
The spectra are the blue curves and the 95\% significance level against the red noise backgrounds are the dashed red curves.
The lower panels present the wavelet spectra with color levels defined by the color bar displayed at the top of Figure~\ref{fig:wt_CME}.}
\label{fig:wt_Flares}
\end{figure}

The frequency analysis applied to the monthly occurrence rate of different classes of flares (A, B, C, M, X) indicates that the most stable and significant oscillations are only seen in the M- and X-classes, consistent with the conclusion of \cite{Gao2016} on the dependence of the occurrence rate and periodicity on flare class.
The periodograms of the M- and X-classes displayed in the top panels of Figure~\ref{fig:wt_Flares} reveal only two periods, 2.1 and 4.4 months for the M-class and none for the X-class.
The global wavelet spectra reveals additional periods, 8.9 months for the M-class, and 4.6 and 7.5 months for the X-class.
The wavelet spectra of the M- and X-classes flare are presented in the bottom panels of Figure~\ref{fig:wt_Flares} and indicate that the periodic patterns were prominent during SC 23 and nearly absent during SC 24.
We note that the 154-day ``Rieger'' periodicity originally found in flares \citep{Rieger1984} and observed during SC 19 to 21 was absent during
SC 23 and 24, the closest value being $\approx$140 days (4.6 months) for the M-class flares.
As mentioned in the above sub-Section, its disappearance took place in SC 22 and persisted during SC 23 \citep{Bai2003} consistent with our result.

%---------------------------------------------------------------------------------------------------
\subsection{Periodicities in Solar Prominences} \label{Per_Protu}
%---------------------------------------------------------------------------------------------------

\begin{figure}[htpb!]
\vspace{0.5cm}
\noindent
\centering
\includegraphics[height=0.93\textwidth,width=0.75\textheight,angle=90]{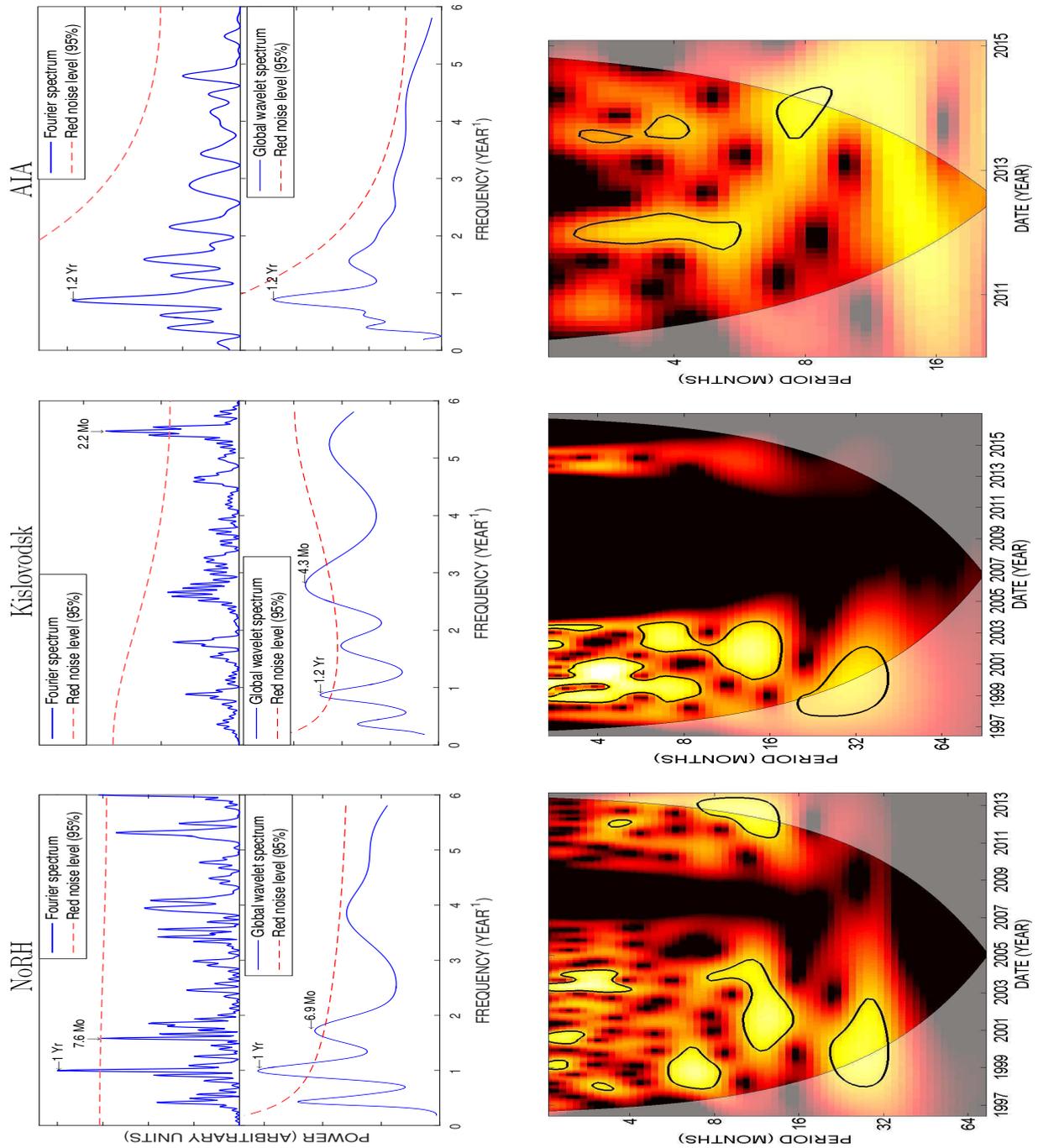}
\caption{Periodicities in the monthly occurrence rate of eruptive prominences.
The left column corresponds to the NoRH data, the central column to the Kislovodsk data, and the right column to the AIA data.
The upper panels present the periodograms (upper sections) and the global wavelet spectra (lower sections) in blue and the 95\% significance level against the red noise backgrounds are shown by dashed red curves.
The lower panels present the wavelet spectra with color levels defined by the color bar displayed at the top of Figure~\ref{fig:wt_CME}.}
\label{fig:periodograms_Protu}
\end{figure}

The frequency analysis applied to the monthly occurrence rates of prominences yields three periods that satisfy the red noise criterion, 7.6 months and 1 year in the NoRH dataset and 2.2 months in the Kislovodsk dataset (Figure~\ref{fig:periodograms_Protu}).
The global wavelet spectra confirm the 1-year period and indicate a 6.9-month (close to the 7.6-month) period in the NoRH dataset, whereas they do not confirm the 2.2-month period in the Kislovodsk dataset and uncover two new periods of 4.3 months and 1.2 year.
The NoRH observations are known to be strongly affected by seasonal (weather) effects, in particular a yearly variation \citep{Shimojo2013}.
It is therefore unclear whether all or part of this 1-year period are of meteorological origin and whether it still preserves a solar origin.
It is interesting to note that the Kislovodsk and AIA datasets show both marked peaks at a close period of 1.2 year although it does not meet the red noise criterion in the latter case.
Weather effects are totally uncorrelated between NoRH and Kislovodsk. The transmission of the SDO data is affected by interruptions due to solar eclipses but they are too short to impact the statistics.

%===================================================================================================
\section{Discussion} \label{Disc}
%===================================================================================================

Our analysis of periodicities in CMEs, proxies of solar activity, flares and prominences reveals an extremely complex situation which, in our opinion, have two different causes: i) the intrinsic difficulty of analyzing non-stationary processes with the consequence that different methods may produce different results, and ii) the complexity of the underlying physical processes.
In order to synthesize the results, facilitate their inter-comparison and identify the main trends, we present below a summary table and several graphs.

\begin{figure}[htpb!]
\vspace{2.01cm}
\noindent
\centering
\includegraphics[width=0.67\textwidth,height=0.77\textheight]{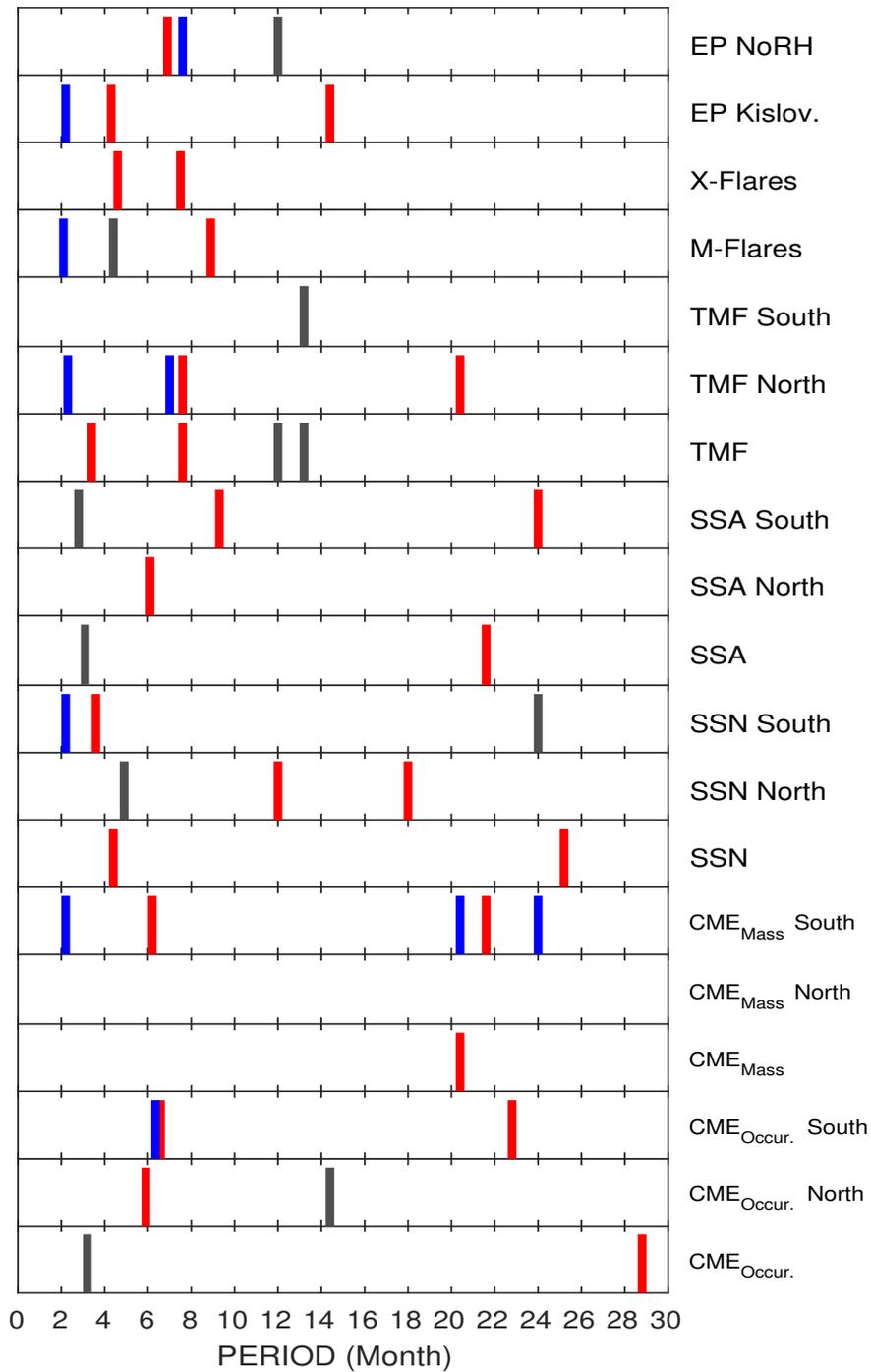}
\caption{Graphical representation of the periods found in the different groups of CMEs, proxies, flares, and eruptive prominences in Fourier spectra (blue bars), global wavelet spectra (red bars) and simultaneously in both spectra (gray bars). The results are for the whole [1996--2016] time interval.}
\label{fig:fourier_SYNTHESE}
\end{figure}

Table~\ref{tab:p_summary} lists the periods detected by the Fourier and global wavelet spectra separately in three ranges and Figure~\ref{fig:fourier_SYNTHESE} presents them in graphical form: for each class of events, the vertical bars indicate the detected periods coded in three colors, blue (Fourier spectra), red (global wavelet spectra), and gray (both spectra).

\begin{table}[htpb!]
  \centering
\caption{Summary of the periods detected over the [1996--2016] time interval with a significance level of 95\% against the red noise backgrounds for CMEs, solar proxies, flares and prominences.
The results of the Fourier analysis are in bold and those of the global wavelet spectra are in italic.}
\vspace{0.2cm}
\label{tab:p_summary}
\begin{footnotesize}
\begin{tabular}{lcccc}
  \hline
                   & 2 Mo -- 11 Mo                                          & 1Yr -- 1.4Yr              & 1.5Yr -- 2.5 Yr                           \\
  \hline

  CME number       & \textbf{\textit{3.2}} Mo                             & --                          &   \textit{2.4} Yr                           \\
  CME number North & \textit{5.9} Mo                                        & \textbf{\textit{1.2}} Yr  & --                                        \\
  CME number South & \textbf{6.4}, \textit{6.5} Mo                        & --                      & \textit{1.9} Yr                           \\

\\

  CME mass         & --                                                                             & --                       &  \textit{1.7} Yr                          \\
  CME mass North   & --                                                   & --                       &  --                                       \\
  CME mass South   & \textbf{2.2}, \textit{6.2} Mo                        & --                                  & \textbf{1.7}, \textit{1.8}, \textbf{2} Yr \\

\\

  SSN              &  \textit{4.4} Mo                                                   & --                       & \textit{2.1} Yr                           \\
  SSN North        &  \textbf{\textit{4.9}} Mo                            & \textit{1} Yr            & \textit{1.5} Yr                           \\
  SSN South        &  \textbf{2.2}, \textit{3.6} Mo                       & --                       & \textbf{\textit{2}} Yr                    \\

\\

  SSA              &  \textbf{\textit{3.1}} Mo                            & --                       & \textit{1.8} Yr                           \\
  SSA North        &  \textit{6.1} Mo                                     & --                       & --                                        \\
  SSA South        &  \textbf{\textit{2.8}}, \textit{9.3} Mo              & --                       & \textit{2} Yr                             \\

\\

  TMF              & \textit{3.4, 7.6} Mo                                 & \textbf{\textit{1, 1.1}} Yr   & --                                        \\
  TMF North        & \textbf{2.3, 7}, \textit{7.6} Mo                           & --                       & \textit{1.7} Yr                           \\
  TMF South        &         --                                           & \textbf{\textit{1.1}} Yr & --                                        \\

\\

  M-class flares   & \textbf{2.1}, \textbf{\textit{4.4}}, \textit{8.9} Mo & --                       & --                                        \\
  X-class flares   & \textit{4.6, 7.5} Mo                                 & --                       & --                                        \\

\\

 EP NoRH           & \textit{6.9}, \textbf{7.6} Mo                        &  \textbf{\textit{1}} Yr  & --                                        \\
 EP Kislov.        & \textbf{2.2}, \textit{4.3} Mo                        &  \textit{1.2} Yr         & --                                        \\
 EP AIA            &         --                                           &  --                      & --                                        \\

 \hline

\end{tabular}
\end{footnotesize}
\end{table}

This figure offers a practical visual overview of the results allowing an easy qualitative comparison.
One sees very limited commonality for periods of less than one year.
The few exceptions are the periods of 3.1--3.2 months found in the occurrence rate of the global set of CMEs and in SSA (and marginally, the 3.4-month period seen in TMF) and those of 5.9--6.1 months found in the CME$_\textrm{N}$ subgroup and in SSA$_{north}$.
It remains however puzzling that the hemispheric results for CMEs and SSA are not fully consistent since CME$_\textrm{S}$ and SSA$_{south}$ exhibits completely different periods, 6.4 and 6.5 months in the first (CME$_\textrm{S}$) case and 2.8 and 9.3 months in the second (SSA$_{south}$) case.
Periods of 1 to 1.2 years are found in the CME$_\textrm{N}$ subgroup, in SSN$_{north}$, in TMF, in TMF$_{south}$ and in the prominence datasets of both NoRH and Kislovodsk (with however a possible bias) but are absent in all other cases.
The same situation roughly prevails for periods clustering around two years observed in CME, CME$_\textrm{S}$, CME$_\textrm{m,S}$, SSN, SSN$_{south}$, SSA, and SSA$_{south}$ and absent otherwise, suggesting that they may be restricted to southern activity.
Note that in the above comparison, we combine periods from the Fourier and global wavelet spectra.
Had we separated them, the commonality would be even more restricted.

\begin{figure}[htpb!]
\noindent
\centering
\includegraphics[width=\textwidth]{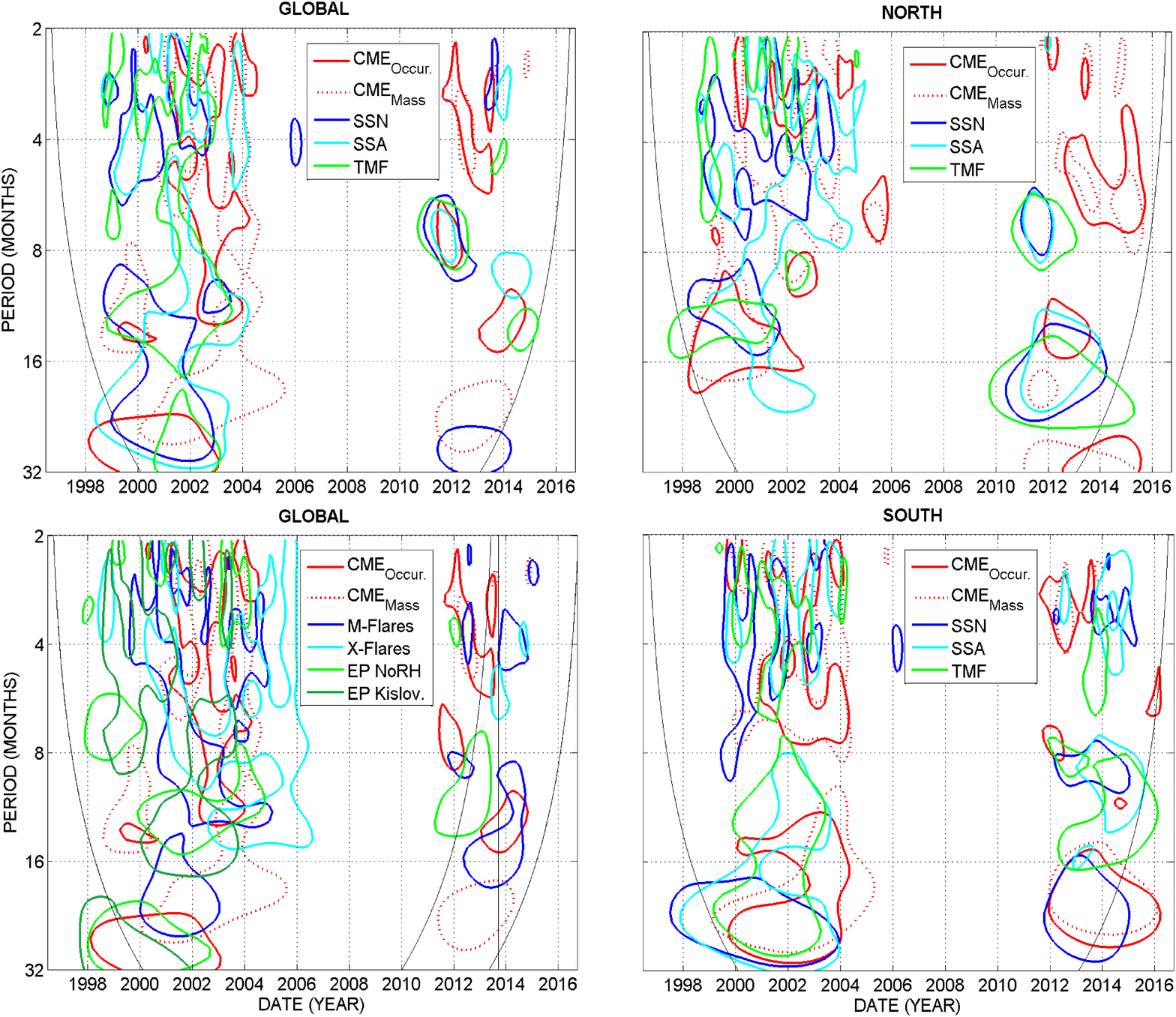}
\caption{Synthetic wavelet spectra where contours corresponding to significance levels of 95\% against the red noise backgrounds for CMEs, solar proxies, flares and prominences are superimposed. Upper left panel: global  CME and solar proxies datasets, upper right panel: CME and solar proxies datasets for the northern hemisphere, lower right panel: CME and solar proxies datasets for the southern hemisphere, and lower left panel: CMEs, flares, and prominences datasets. In this latter case, there are distinct cones of influences resulting from the limited time interval of the AIA prominences data.}
\label{fig:wt_SYNTHESE}
\end{figure}

Figure~\ref{fig:wt_SYNTHESE} synthesizes the different wavelet spectra generated in the previous sections.
Closed contours delimiting regions of statistically significant signal defined as exceeding the 95\% significance level against the red noise backgrounds are superimposed for the CMEs and the proxies (globally and by hemisphere) on the one hand and for CMEs, flares and eruptive prominences on the other hand.
The four panels conspicuously show that for all considered processes, periodicities are restricted to the maxima of solar activity and are absent during the two minima of the 1996--2016 time interval.
In agreement with the frequency analysis, the contours broadly cluster into three time domains.
For periods less than one year, they appear as a set of ``islands'' with very partial overlap between the different processes.
The clustering is even less pronounced in SC 24 with a very limited number of islands.
Note however an exception with a remarkable superposition of the contours for CME occurrence rate, SSN, SSA, and TMF that took place around 2012 with periods ranging from $\approx$6 to $\approx$9 months.
At larger periods exceeding 9 months, the contours are better defined and are distributed in two broad regions centered at periods of $\approx$1 and $\approx$2 years.
But there is a clear dichotomy between these two regimes best illustrated in SC 23 as the $\approx$1-year period is only present in the northern hemisphere whereas the $\approx$2-year period is only present in the southern one; note the good superposition of the contours for CMEs and the three proxies in both cases.
This trend reappears in SC 24 with however a subtle difference as the one-year period shifts to $\approx$1.4 year in the northern hemisphere whereas it stays at $\approx$2 years in the southern one.
As a matter of fact and as expected, the global behaviours simply reflect the combined oscillation patterns that take place separately in the two hemispheres.
The situation is far less clear when comparing CMEs, flares and prominences with extremely limited superposition of contours.
Indeed, the only noteworthy exception is the $\approx$2-year period found in both CME occurrence and mass rates and in M-flares and prominences during the ascending phase of SC 23.

\begin{figure}[htpb!]
\noindent
\centering
\includegraphics[width=\columnwidth]{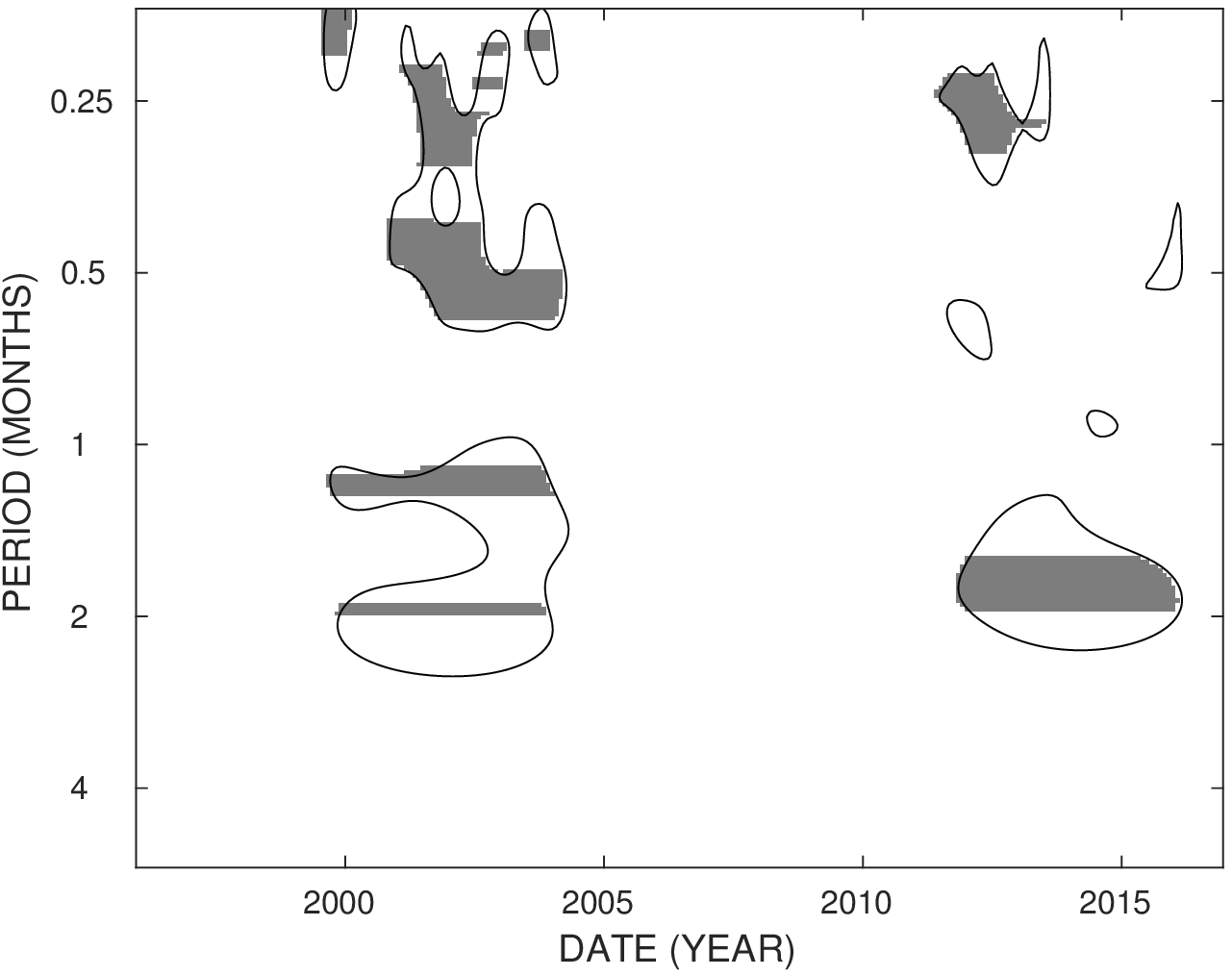}
\caption{Illustration of the method of estimating the stability of the periodic regimes in the wavelet spectra.}
\label{fig:period}
\end{figure}

As it is difficult to appreciate the stability of the various periodic regimes on the wavelet spectra, that is how long the periodic patterns extend, we attempt to quantify the duration of these regimes by considering (horizontal) slices in the wavelet spectra with a width of one month.
Whenever a slice intersects a closed contour, we measure the length of the intercepted section as illustrated in Figure~\ref{fig:period} and compare it to the period $P$ corresponding to that slice.

\begin{figure}[htpb!]
  \centering
  ~~~~~Two-period~stability \hspace{2.2cm} ~~~~~~Four-period~stability
  \includegraphics[height=0.81\textheight,width = 0.47\textwidth]{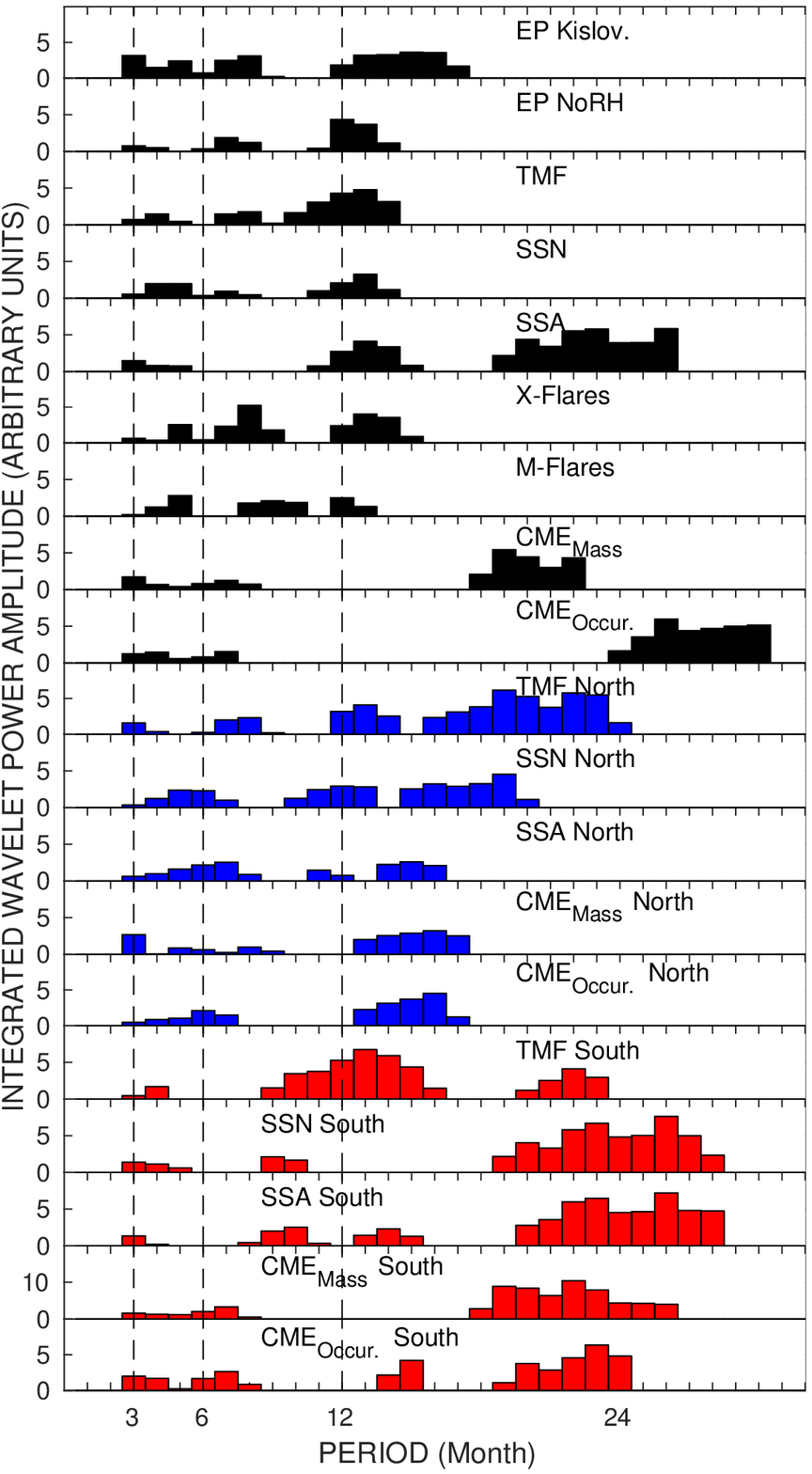}
  ~~~\includegraphics[height=0.81\textheight,width = 0.47\textwidth]{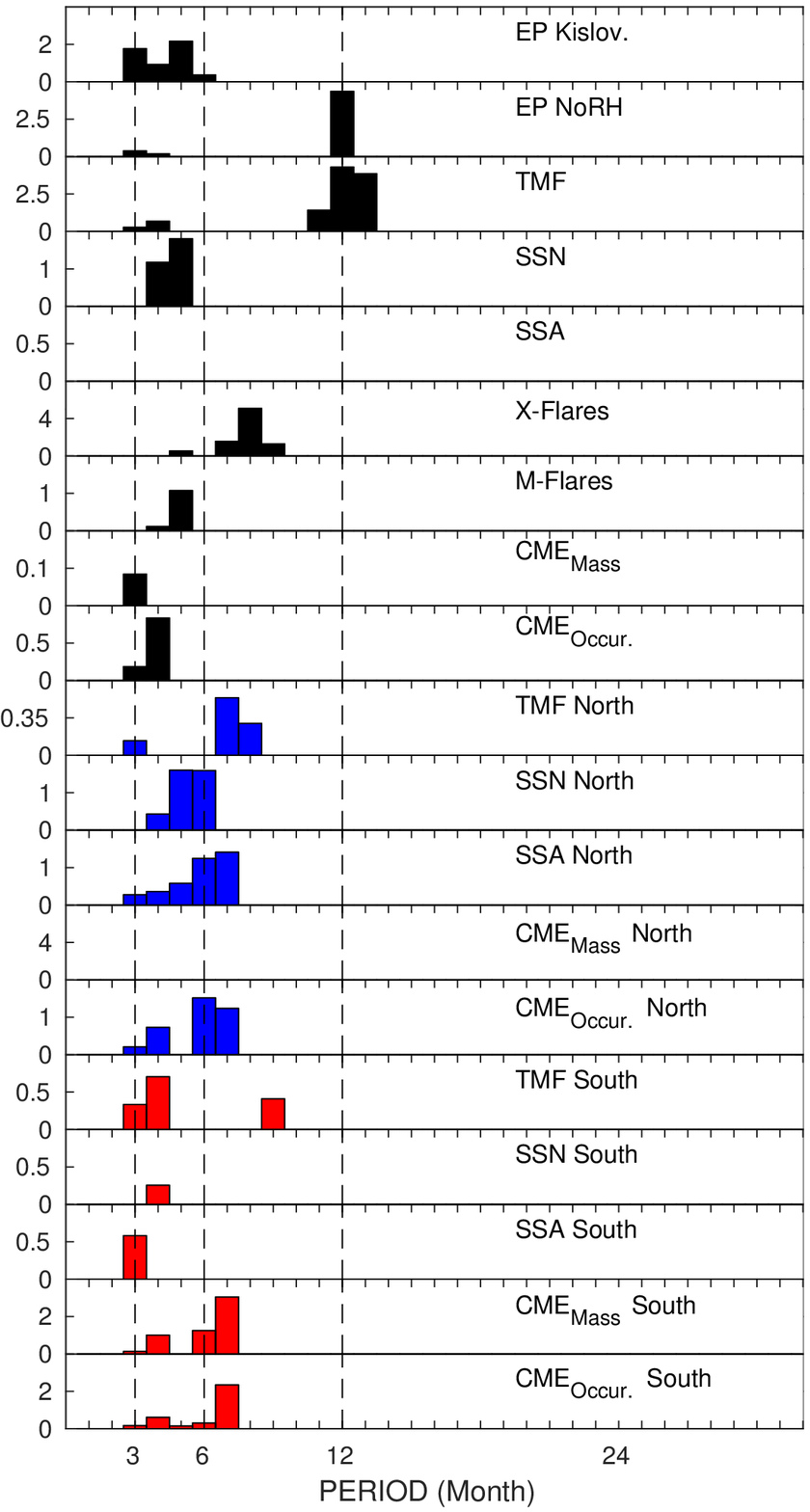}
  \caption{Distributions of periods found in CME rates, solar proxies, M- and X-class flares and prominences based on the time-frequency (wavelet) analysis. Two cases of stability are displayed depending upon the number of observed periods: two periods (left panel) and four periods (right panel). The vertical dashed lines at 3, 6, and 12 months are intended to guide the eyes. The results in the two hemispheres are emphasized by different colors: blue (north) and red (south).}
\label{fig:p_SYNTHESE}
\end{figure}

Whenever this length exceeds a preset integer number $k$ of the period $P$, we integrate the spectral power in the relevant section.
We finally build the distribution of integrated power as a function of period with therefore a time resolution of one month.
We consider time intervals corresponding to $k$=2, 3 and 4 but only display the results for the two extreme values in Figure~\ref{fig:p_SYNTHESE}.
In the case of the weak constraint $k$=2, we see that, with a few exceptions, all considered processes exhibit high- and mid-frequency oscillations with periods of $\approx$3 months and $\approx$1 year whereas the period of $\approx$6 months is less frequent.
But most striking is the dichotomy between the two hemispheres as already emphasized above, with periods of 1--1.5 years prevailing in the northern hemisphere whereas a period of $\approx$2 years prevails in the southern one.
The only notable exception is TMF which, in particular, exhibits a marked one-year period in the southern hemisphere.
The more stringent condition $k$=4 confirms that the $\approx$3-month oscillation is present in most -- but not all -- processes whereas the $\approx$6-month period is present in only a few.
The one-year period persists in only the TMF and the prominences, and the two-year period has disappeared simply because the condition $k$=4 translates in a duration of 8 years, largely over what we observe.
It should be kept in mind that the temporal information is lost in Figure~\ref{fig:p_SYNTHESE}, in particular the differences between the two Solar Cycles so that this figure should be interpreted in the context of Figure~\ref{fig:wt_SYNTHESE}.

Period searching is notoriously difficult and this is the reason why many different methods have been developed and a safe approach is to implement several of them to ascertain the robustness of the results.
It is even more complex in the case of non-stationary processes and putative periodic patterns may change with time or even disappear and this is certainly the case of solar phenomena as this is becoming more and more obvious as observations accumulate.
The complexity of the solar processes and their interplay further add to the difficulty of assessing the periodicities and interpreting them.
We do find that both CME occurrence and mass rates exhibit statistically significant oscillations with short- and mid-range periods but, out of the dozen periodicities exceeding two months in the occurrence rate indicated by spectral analysis in the past studies of \cite{Lou2003}, \cite{Lara2008}, and \cite{Choudhary2014}, we confirm only one, namely 3.2 months.
We do find a few others, most notably the 6.2--6.4 months, but only in restricted datasets, either in time or in hemisphere (it may be an integral multiple of the above period, namely twice 3.2 months).
Part of the problem may stem from the use by past studies of the CDAW data for the occurrence rate and from its large deviation from the solar cycle variation (see for instance Figure 1 of \cite{Choudhary2014}) contrary to the ARTEMIS and other catalogs (see Figure 8 of \cite{Lamy2014}).
As a matter of fact, our in-depth analysis uncovers a situation far more complex than found in these past studies (and indeed, far more complex than we had anticipated) where CME rates depend upon both solar cycle (and even phase in the SC) and hemisphere although \cite{Choudhary2014} did observe a change of period in the CME occurrence rate, namely from 182$\pm$1 days ($\approx$6 months) during the maximum of SC 23 to $\approx$154 days ($\approx$5 months) during that of SC 24.
On the positive side, it shows that the CME activity exhibits statistically significant oscillations with short- and mid-range periods whose properties are common to all solar, coronal, and heliospheric processes: variable periodicity, intermittency, asymmetric development in the northern and southern solar hemispheres, and largest amplitudes during the maximum phase of solar cycles \citep{Bazilevskaya2014}.

Whereas the strict localization of periods from our spectral analysis indicate little commonality between CME and solar proxies, our wavelet spectra reveal conspicuous broad regions where the contours delimiting regions of statistically significant signal for the different CME datasets, SSN, SSA, and TMF nicely overlap, strongly suggesting common underlying processes.
Surprisingly, \cite{Choudhary2014} have not obtained a similar result when comparing CME and SSA counts.
However, we agree with them on the lack of commonality between CMEs and flares and further emphasize that we do not observe the ``Rieger'' periodicity in flares which has prevailed during past solar cycles.
\cite{Lou2003} have argued that, whereas the CME count is nearly complete, that of flares is not because only half of the Sun is observed at any given time (likewise, we may remark that eruptive prominences are observed as limb events).
But their statistical test indicates that a more complete flare dataset would not affect the quasi-periodic features in the power spectrum.
A more plausible explanation lies in the fact that the general terminology of ``coronal mass ejection'' hides a diversity of phenomena that are triggered by either previous accumulation of magnetic energy through flux emergence and foot-point motions, magnetic field reconnection at coronal heights, or flux-rope (filament) eruption to name the most prominent processes.
These processes may have different periodicities which would be blurred when considering CMEs as an homogeneous population.
Probably this could be disentangled if CMEs were easily separated on the basis of their origin (a very difficult task presently limited to a small number of events).
However, there may be a more profound reason, at least in the case of CMEs and flares as they require different magnetic configurations to trigger an explosion \citep{Choudhary2014}.
Whatever the case, we consider our result of conspicuous commonality of periodic patterns between CMEs and solar proxies based on wavelet spectra to be highly significant and linking these patterns to underlying periodicities in the emerging magnetic flux which are thought to be related to the dynamics of the deep layers of the Sun \citep{Rieger1984,McIntosh2015} and intrinsic to the solar dynamo mechanism.
Prominent periods are then generated by stochastic processes caused by the periodic emergence of magnetic flux as the solar cycle progresses, as proposed by \cite{Wang2003}, who also pointed out that there is no reason for a pattern of stable and reproducible periods as indeed observed.
Alternatively, the periodic pattern may be imprinted to the emerging magnetic field by equatorially trapped Rossby-type waves at and beneath the solar photosphere \citep{Lou2003}.

%===================================================================================================
\section{Conclusion} \label{Concl}
%===================================================================================================

Our analysis of the CME occurrence and mass rates based on the ARTEMIS II catalog of LASCO-C2 observations over 21 years [1996 -- 2016], of the occurrence rates of solar
flares and prominences, and of solar activity proxies globally and by hemispheres further considering various phases of SC 23 and 24 allows us to make the main conclusions summarized below.

\begin{enumerate}
    \item
Many periods ranging from 2.1 months to 2.4 years are found among the 19 classes of events that we have introduced but none are common to the whole set of classes
(Figure~\ref{fig:fourier_SYNTHESE}).
The shortest periods of 2.1--2.3 months are observed in the CME$_\textrm{m,S}$ subgroup, SSN$_{south}$, TMF$_{north}$, M-flares and the Kislovodsk dataset of prominences.
Periods of 3.1--3.4 months are observed in the occurrence rate of CMEs (a robust determination since the 3.2-month period is yielded by both the Fourier and global wavelet spectra), SSA, and TMF.
Periods of 5.9--6.5 months are observed in the CME$_\textrm{S}$ and CME$_\textrm{m,S}$ subgroups, and SSA$_{north}$.
Periods of 1--1.2 year are observed in CME$_\textrm{N}$ subgroup, SSN$_{north}$, TMF, TMF$_{south}$, the NoRH and Kislovodsk datasets of prominences (in the latter two cases, a terrestrial origin of the one-year period is possible).
Periods of 1.7--2.4 years are observed in occurrence and mass rates of CMEs, in the CME$_\textrm{S}$ and CME$_\textrm{m,S}$ subgroups, SSN, SSN$_{south}$, SSA,
SSA$_{south}$, and TMF$_{north}$.

    \item
Whereas the short-range periods of $\approx$3 and $\approx$6 months are present in both solar
hemispheres, the mid-range periods of $\approx$1 and $\approx$2 years exhibit a very distinct
behaviour with the first one ($\approx$1 year) being present only in the northern hemisphere
and the second one ($\approx$2 years) only in the southern hemisphere with however a few
exceptions.

    \item
The observed periodic activity shares all properties common to all solar, coronal, and heliospheric processes: variable periodicity, intermittency, asymmetric development in the northern and southern solar hemispheres, and largest amplitudes during the maximum phase of the Solar Cycles.
This is particularly the case of the periodic behaviour of CMEs which prevails during these maximum phases -- an effect even more pronounced in SC 23 than in the weaker SC 24 -- and tends to vanish during the minimum phases.

    \item
Common periodicities shared by CMEs and solar proxies are most likely linked to their direct relationship to the magnetic flux emergence.

\end{enumerate}

%===================================================================================================
\section*{Acknowledgments}
%===================================================================================================
We are grateful to M. Shimojo for providing an updated list of NoRH prominences, and E. Pardo-Ig\'uzquiza for providing his software for the maximum entropy method.
I.~Toth acknowledges the support from project GINOP-2.3.2-15-2016-00003 ``Cosmic effects and hazards''.
The LASCO-C2 project at the Laboratoire d'Astrophysique de Marseille and at Laboratoire Atmosph\`eres, Milieux et Observations Spatiales is funded by the Centre National d'Etudes Spatiales (CNES).
LASCO was built by a consortium of the Naval Research Laboratory, USA, the Laboratoire d'Astrophysique de Marseille (formerly Laboratoire d'Astronomie Spatiale), France, the Max-Planck-Institut f\"ur Sonnensystemforschung (formerly Max Planck Institute f\"ur Aeronomie), Germany, and the School of Physics and Astronomy, University of Birmingham, UK.
 The primary data supporting the analysis are available at http://lasco-www.nrl.navy.mil/.
The secondary data produced at the Laboratoire d'Astrophysique de Marseille and at the Laboratoire Atmosph\`eres, Milieux et Observations Spatiales are available from the third author upon request and subject to an agreement.
SOHO is a project of international cooperation between ESA and NASA.

%===================================================================================================

\end{document}